\documentclass[12pt]{iopart}

\usepackage{float}
\usepackage{epsfig}
\usepackage{dsfont}
\usepackage{amsmath}
\usepackage{braket}
\usepackage{amsfonts}
\usepackage{amssymb}
\usepackage{marginnote}
\usepackage{color,soul}
\usepackage{mathtools}
\usepackage{graphicx}  
\usepackage[dvipsnames]{xcolor}
\usepackage{todonotes}
\usepackage{bbm}
\usepackage{tikz}
\usepackage{multirow}
\usepackage{lmodern}
\usepackage{braket}
\usepackage{wrapfig}
\usepackage{comment}
\newcommand\newblock{\hskip .11em\@plus.33em\@minus.07em}
\RequirePackage[colorlinks,citecolor=blue,urlcolor=magenta,linkcolor=blue]{hyperref}
\usepackage[square,numbers,sort&compress]{natbib}
\usepackage[left=2.7cm,right=2.7cm,bottom=2.75cm]{geometry}

\begin{document}

\title[Imprints of the operator ordering ambiguity on the dynamics of perfect fluid \dots]{Imprints of the operator ordering ambiguity on the dynamics of perfect fluid dominated quantum universe}

	\author{Harkirat Singh Sahota}
	
	\address{Department of Physical Sciences, Indian Institute of Science Education \& Research (IISER) Mohali, Sector 81 SAS Nagar, Manauli PO 140306 Punjab India.}
\ead{harkirat221@gmail.com, ph17078@iisermohali.ac.in}
    \vspace{10pt}
\begin{indented}
        \item[] July 2024
\end{indented}
    \begin{abstract}
    Sharply peaked quantum states are conjectured to be conducive to the notion of a quantum-corrected spacetime. We investigate this conjecture for a flat-FLRW model with perfect fluid, where a generalized ordering scheme is considered for the gravitational Hamiltonian. We study the implications of different ordering choices on the dynamics of the quantum universe. We demonstrate that the imprints of the operator ordering ambiguity are minimal, and quantum fluctuations are small in the case of sharply peaked states, leading to a consistent notion of a quantum-corrected spacetime defined via the expectation value of the scale factor. Surprisingly, the ordering imprints survive far away from the singularity through the quantum fluctuations in the quantum-corrected spacetime for broadly peaked states.
    \end{abstract}
	
\section{Introduction}
    The operator ordering ambiguity is one of the many issues at the heart of all canonical approaches to quantizing gravity \cite{kiefer_quantum_2012}. Ordering ambiguity in quantum gravity appears in two avatars: the structure of the Hamiltonian constraint involves the product of non-commuting variables leading to inequivalent constraint operator choices \cite{dewitt_quantum_1967}, and implementation of constraint algebra at the operator level leading to quantum anomalies \cite{Tsamis_ordering}. In this work, we will address the first case in the context of a quantum mechanical model of gravity, i.e., a minisuperspace system with finite degrees of freedom.  Even in the early seminal work of DeWitt \cite{dewitt_quantum_1967}, the ordering prescription is proposed for the Hamiltonian constraint, the Laplace-Beltrami ordering, based on physical arguments pertaining to the covariance of the differential operator in the space of 3-geometries \cite{Halliwell_ordering}. However, there is still no consensus on the preferred ordering for the Hamiltonian constraint, and several other choices are also proposed \cite{Hawking:1985bk,Vilenkin_ordering}.
 
    The analyses that consider a generalized scheme of the ordering of the Hamiltonian constraint are hard to find; one that deals with how the ordering choice affects the evolution of the quantum universe is further obscure. Still, there are a handful of analyses that address the operator ordering ambiguity at some level, e.g., see \cite{Louko:1988zb,Rostami_ordering,Halliwell,Kontoleon:1998pw,Bojowald:2002gz,Bojowald:2014ija,kiefer_singularity_2019,Matsui:2021yte,Sahota2023,Sahota_Infrared}. A typical quantum gravity analysis deals with the status of a singularity \cite{Thebault:2022dmv}, leading to the notion of a quantum-corrected spacetime, with the understanding that quantum effects are relevant only near the singularity \cite{Bojowald:2007,Ashtekar:2009mb,Agullo:2012fc,Agullo:2012sh,Agullo:2013ai}. However, it is an important (although tedious) exercise to demonstrate these notions to be ordering independent, i.e., the ordering imprints do not leak into the semiclassical regime via the quantum-corrected spacetime. %The irrelevance of operator ordering in the WKB analysis is argued in various works, e.g., see \cite{Louko:1988zb,Kiefer:1993fg,Bojowald:2014ija}, 
    The focus of this work is on the identifications of states for which the dynamics of the quantum universe is agnostic to the ordering chosen. As it happens, such states are of importance in the discussion surrounding a consistent notion of quantum-corrected spacetime defined via the expectation value of the metric variables \cite{Ashtekar:2009mb,Sahota2023}. 

    The notion of a quantum-corrected spacetime is conjectured to be well defined for a state that is sharply peaked on the classical trajectory away from the singularity, and near-singularity it is peaked on an effective geometry undergoing a quantum bounce \cite{Ashtekar:2009mb}. The moments of the scale factor appearing in the perturbation Hamiltonian effectively capture the quantum fluctuations, and one can introduce a quantum-corrected geometry through the expectation value of the scale factor, leading to a consistent semiclassical analysis. In \cite{Sahota2023}, we have demonstrated the consistency of such a simplification, where the consideration of a sharply peaked state turned out to be the crucial assumption. The expectation value of geometric quantities is shown to match the quantities computed from the expectation value of the metric variables in the leading order in the parameter that determines the shape of the distribution, thus verifying the conjecture in \cite{Ashtekar:2009mb}. The question we would like to address in this analysis is; {\it Do the ordering choice leave any imprint on the quantum-corrected spacetime?} 
    
    In this work, we investigate the imprints of ordering ambiguity on the dynamics of the quantized flat-FLRW universe with perfect fluid. We construct a unitarily evolving wave packet and study the evolution of the probability distribution associated with it for different ordering choices. The expectation value of the scale factor has global non-zero minima at the classical singularity and provides the notion of quantum-corrected spacetime that represents a universe undergoing quantum bounce. The goal of this analysis is to determine how the different ordering choices manifest into the notion of a quantum-corrected spacetime and to study the fluctuations in the quantum-corrected spacetime. We start with the canonical formulation of the model and discuss the classical behavior of the model in Sec. \ref{Sec2}. The model is quantized in Sec. \ref{Sec3}, and we discuss the behavior of the probability distribution associated with the wave packet. In Sec. \ref{Sec5}, we investigate the imprints of ordering choice on the quantum dynamics of the universe and summarize our findings in Sec. \ref{Sec6}.

\section{FLRW Model with Schutz Fluid}\label{Sec2}

\noindent We are interested in the dynamics of a flat-FLRW universe
\begin{align}
    ds^2=-\mathcal{N}^2(\tau)d\tau^2+a^2(\tau)d{\bf x}^2,
\end{align}
with a perfect fluid as matter. The Hamiltonian constraint for this model with Schutz's parameterization of the perfect fluid takes the form \cite{Schutz_1970, Schutz_1971},
\begin{align}
	\mathcal{H}\coloneqq\mathcal{N}\left[-\frac{p_a^2}{2a}+\frac{p^{}_T}{a^{3\omega}}\right]\approx 0,
\end{align}
where $a$ is the scale factor, $\mathcal{N}$ is the lapse function, $p_a$ is momentum conjugate to the scale factor, $T$ is the fluid degree of freedom, $p_T$ is momentum conjugate to the fluid variable, and $\omega$ is the equation of state parameter of the perfect fluid.\footnote{Here, we have rescaled the fluid momentum $p_T$ with the volume of auxiliary cell $V_0$ and used $4\pi G/3V_0=1$.} The lapse function that is compatible with the choice of fluid variable as the clock degree of freedom is $\mathcal{N}=a^{3\omega}$. With this choice, we have $\dot{T}=1$, and the fluid variable is linearly related to the coordinate time $\tau$. The Hamiltonian constraint, in this case, becomes
\begin{align}
	\mathcal{H}=-\frac{1}{2}a^{3\omega-1}p_a^2+p_T\approx 0.\label{HS}
\end{align}
The equations of motion with this gauge choice are
\begin{equation}
	\begin{aligned}
	\dot{T}&=1\qquad\&\qquad\dot{p}^{}_T=0,\\
	\frac{\ddot{a}}{a}&+\frac{1-3\omega}{2}\left(\frac{\dot{a}}{a}\right)^2=0,\qquad\&\qquad a^{1-3\omega}\dot{a}^2=2p_T.
	\end{aligned}
\end{equation}
The momentum conjugate to the fluid variable, $p_T$, is the standard constant of motion for perfect fluid cosmology, $\rho\, a^{3(1+\omega)}$, and is a Dirac observable of the system. The scale factor in this gauge behaves as
\begin{align}
	a(\tau)=\left(\frac{9p_T(1-\omega)^2}{2}\right)^{1/3(1-\omega)}\tau^{2/3(1-\omega)}.\label{sfC}
\end{align}
The solution space of this model consists of two branches, an expanding $(\tau>0)$ and a collapsing $(\tau<0)$ universe and the universe remains in either of these trajectories throughout its evolution. However, the question of initial singularity is still fluid dependent, where the singularity is a curvature singularity for all fluid choices except $\omega=-1$, where the system has a coordinate singularity at $\tau=0$, and the curvature invariants are regular. Another reason to consider the dust dominated universe is the bouncing scenario in dust dominated quantum cosmology leads to the scale-invariant power spectrum for scalar and tensor perturbations \cite{Peter:2008qz}. Therefore, we limit the scope of this analysis to the cases of a singular dust dominated universe and a regular cosmological constant driven universe, where we will present the results for the latter case in the main text and the earlier case in the \ref{App1} and \ref{App2}.

\section{Quantum Model}\label{Sec3}
The aim is to write a generalized ordering for the operator corresponding to the phase space function in Eq. \eqref{HS} and study the imprints of ordering on the dynamics in this quantum model. We will follow the operator representation introduced in \cite{kiefer_singularity_2019} and generalize it for the case of an arbitrary equation of state parameter. The Wheeler-DeWitt equation, in this case, takes the form of Schr\"odinger equation where the fluid variable plays the role of time
\begin{align}
	i\frac{\partial\Psi}{\partial \tau}=\frac{1}{2}a^{3\omega-1+p+q}\frac{d}{da}a^{-p}\frac{d}{da}a^{-q}\Psi.\label{HSO}
\end{align}
The parameters $p$ and $q$ represent the freedom in choosing the ordering, and we are working with $\hbar=1$. The Hamiltonian operator is symmetric on the Hilbert space $L^2(\mathbb{R}^+,a^{1-3\omega-p-2q}da)$ with the inner product
\begin{align}
    \braket{\psi|\chi}=\int_0^\infty da\,\psi^*(a,\tau)\chi(a,\tau)a^{1-3\omega-p-2q}
\end{align}
and the discussion about the self-adjointness of this operator will follow the analysis in \cite{kiefer_singularity_2019}. In this case, the operator is essentially self-adjoint for $|1+p|\geq3(1-\omega)$, and there exists a family of self-adjoint extensions for $|1+p|<3(1-\omega)$, with the boundary condition, in this case, parameterized by an angle $\theta\in[0,2\pi)$. The representation of the momentum operator that is symmetric with the choice of measure is given by
\begin{align}
	\hat{p}_a=-ia^{-\frac{1-3\omega-p-2q}{2}}\frac{d}{da}a^{\frac{1-3\omega-p-2q}{2}}.
\end{align}
The Hamiltonian operator can be written as
\begin{align}
	\hat{\mathcal{H}}_g=-\frac{1}{2}\hat{a}^{\frac{3\omega-1+p}{2}}\hat{p}_a\hat{a}^{-p}\hat{p}_a\hat{a}^{\frac{3\omega-1+p}{2}}.
\end{align}
In this form, the ordering parameter $q$ does not appear in the formalism, which we will show later on is the free parameter of the theory. The solution of the WDW equation \eqref{HSO} is obtained by the separation ansatz $\Psi(a,\tau)=e^{iE\tau}\phi_E(a)$, leading to the eigenvalue equation $\hat{\mathcal{H}}\psi_E=-E\psi_E$. The spectrum of the Hamiltonian operator comprises of negative as well as positive eigenvalues, where the spectrum is continuous for $E>0$ and discrete for $E<0$ \cite{kiefer_singularity_2019}. We restrict the analysis to the positive energy only, and the stationary states with $E>0$ are given by
\begin{align}
	\begin{aligned}
			\phi_E^1(a)&=a^{\frac{1}{2}(1+p+2q)}J_{\frac{|1+p|}{3(1-\omega)}}\left(\frac{2\sqrt{2E}a^{\frac{3}{2}(1-\omega)}}{3(1-\omega)}\right)\\
			\phi_E^2(a)&=a^{\frac{1}{2}(1+p+2q)}Y_{\frac{|1+p|}{3(1-\omega)}}\left(\frac{2\sqrt{2E}a^{\frac{3}{2}(1-\omega)}}{3(1-\omega)}\right).
		\end{aligned}\label{Stast}
\end{align}
In stationary states, the ordering parameter $q$ appears in the exponent only, and the $q$-dependence of the measure cancels it in physical constructs of the theory, such as the probability distribution $a^{1-3\omega-p-2q}|\psi|^2$ or the expectation values of observables computed from the states constructed from the stationary states. The ordering imprints survive only through the $p$-dependence coming from the index of the Bessel function, as the exponents are canceled for both $p$ and $q$ dependence. For some phase space observable $O=a^np_a^m$, symmetrized trivially as $\hat{O}=\hat{a}^{n/2}\hat{p}_a^m\hat{a}^{n/2}$, the expectation value in a wave packet $\displaystyle\psi(a,\tau)=\int_0^\infty A(E)\phi_E(a)e^{iE\tau}=a^{(1+p+2q)/2}\chi(a,\tau,p)$ takes the form

\begin{align}
    \braket{\psi|\hat{O}|\psi}=&\int_0^\infty da\;a^{1-3\omega-p-2q}\psi^*(a,\tau)\hat{a}^{n/2}\hat{p}_a^m\hat{a}^{n/2}\psi(a,\tau)\nonumber\\
    =&(-i)^m\int_0^\infty da\;a^{1-3\omega-p-2q}a^{(1+p+2q)/2}\chi^*(a,\tau,p)a^{n/2}a^{-(1-3\omega-p-2q)/2}\nonumber\\
    &\qquad\qquad\qquad\frac{\partial^m}{\partial a^m}\left[a^{(1-3\omega-p-2q)/2}a^{n/2}a^{(1+p+2q)/2}\chi(a,\tau,p)\right]\nonumber\\
    =&(-i)^m\int_0^\infty da\;a^{(2-3\omega+n)/2}\chi^*(a,\tau,p)\frac{\partial^m}{\partial a^m}\left[a^{(2-3\omega+n)/2}\chi(a,\tau,p)\right],
\end{align}
which is independent of parameter $q$. This result can be shown for other ordering choices of the phase space observable as well. To facilitate the construction of unitarily evolving wave packets, we require orthonormal stationary states. Owing to the orthogonality of Bessel functions \cite{ArfkenWeber},
\begin{align}
	\int_0^\infty x\;J_\nu(\lambda x)J_\nu(\lambda' x)dx=\frac{\delta(\lambda-\lambda')}{\lambda},\quad\text{for }\nu>-\frac{1}{2},
\end{align}\\
\noindent the stationary states $\phi_E^1$ form an orthogonal set in the Hilbert space
\begin{align}
	\braket{\phi_E^1|\phi_{E'}^1}=\frac{3(1-\omega)}{4\sqrt{E}}\delta(\sqrt{E}-\sqrt{E'}).
\end{align}
For these stationary states, the self-adjoint extension $\theta=\pi$ is appropriate, and the Hamiltonian is self-adjoint for all ordering choices \cite{kiefer_singularity_2019}. In the limit $x\rightarrow 0$, $J_\nu(x)$ has the asymptotic form
\begin{align}
    J_\nu(x)\sim \frac{1}{\Gamma(\nu+1)}\left(\frac{z}{2}\right)^\nu,\quad\text{where }\nu\neq-1,-2,-3,\dots 
\end{align}
The behavior of the probability distribution for these positive energy stationary states $\phi_E^1$ near the singularity is
\begin{align}
	\lim_{a\rightarrow0}a^{1-3\omega-p-2q}|\phi^1_E(a)|^2=&\lim_{a\rightarrow0}a^{1-3\omega-p-2q}a^{1+p+2q}\left(\frac{2\sqrt{2E}a^{3(1-\omega)/2}}{3(1-\omega)}\right)^{2\frac{|1+p|}{3(1-\omega)}}\nonumber\\
    \sim&\lim_{a\rightarrow 0}a^{2-3\omega+|1+p|}.
\end{align}
Therefore, the probability distribution vanishes provided that $2-3\omega+|1+p|>0$, which is the case for $\omega\leq2/3$, i.e., the singularity is resolved for the cases that we are interested in, dust dominated and cosmological constant driven universe. We will construct the wave packet from these states using a Poisson-like energy distribution 
\begin{align}	
    \psi(a,\tau)&=\int_0^{\infty}d\sqrt{E}A(\sqrt{E})\tilde{\phi}_E^1(a)e^{iE\tau}\\
	A(\sqrt{E})&=\frac{\sqrt{2}\lambda^{\frac{1}{2}(\kappa+1)}}{\sqrt{\Gamma(\kappa+1)}}\sqrt{E}^{\kappa+\frac{1}{2}}e^{-\frac{\lambda}{2}E},\label{dis}
\end{align}
following the choice made in \cite{hajicek_singularity_2001} and here $\tilde{\phi}^1_E(a)$ is the normalized stationary state. The allowed parameter regime is $\kappa>-1$ and $\lambda>0$. The mean and width of this distribution are
\begin{align}
	\overline{E}=\frac{\kappa+1}{\lambda},\quad\frac{\Delta E}{\overline{E}}=\frac{1}{\sqrt{\kappa+1}}.\label{disP}
\end{align}
Here, the stationary states are labeled by the eigenvalue of the operator $\hat{p}_T$, whose classical counterpart identifies the different trajectories in the solution space, as seen in the Eq. \eqref{sfC}. In this treatment of quantum gravity, the statistical interpretation of quantum mechanics \cite{Ballentine} is appropriate. The classical counterpart to the quantum system that the wave packet represents is, in fact, the ensemble of universes. Here, the probability distribution for the statistical system is given by the energy distribution \eqref{dis} used to construct the wave packet. Therefore, the classical properties of the system are the ensemble averages of the on-shell expressions of the geometric quantities of interest. Therefore, in this picture, a sharply peaked state is identified with the energy distribution of vanishing width, i.e., $\kappa\rightarrow\infty$. The analytical expression for the wave packet takes the form of Kummer's confluent hypergeometric function, $\;_1F_1(a; b; z)$ \cite{GRADSHTEYN1980635}

	\begin{align}
			\psi(a,\tau)=\sqrt{\frac{3(1-\omega)}{\Gamma\left(\kappa+1\right)}}&\left(\frac{\sqrt{2}}{3 (1-\omega )}\right)^{\frac{| p+1| }{3 (1-\omega )}+1}\lambda ^{\frac{\kappa +1}{2}} \left(\frac{\lambda }{2}-i \tau \right)^{-\frac{| p+1| }{6 (1-\omega )}-\frac{\kappa }{2}-1}\frac{\Gamma \left(\frac{\kappa }{2}+\frac{| p+1| }{6 (1-\omega )}+1\right) }{\Gamma \left(\frac{| p+1| }{3 (1-\omega )}+1\right) } \nonumber\\
			&\hspace{-2cm}a^{\frac{1}{2} (| p+1| +p+2 q+1)}\, _1F_1\left(\frac{\kappa }{2}+\frac{| p+1| }{6 (1-\omega )}+1;\frac{| p+1| }{3 (1-\omega )}+1;-\frac{2 a^{3 (1-\omega )}}{9 (1-\omega )^2 \left(\frac{\lambda }{2}-i \tau \right)}\right).\label{WP}
		\end{align}

\begin{figure}
        \includegraphics[width=\textwidth]{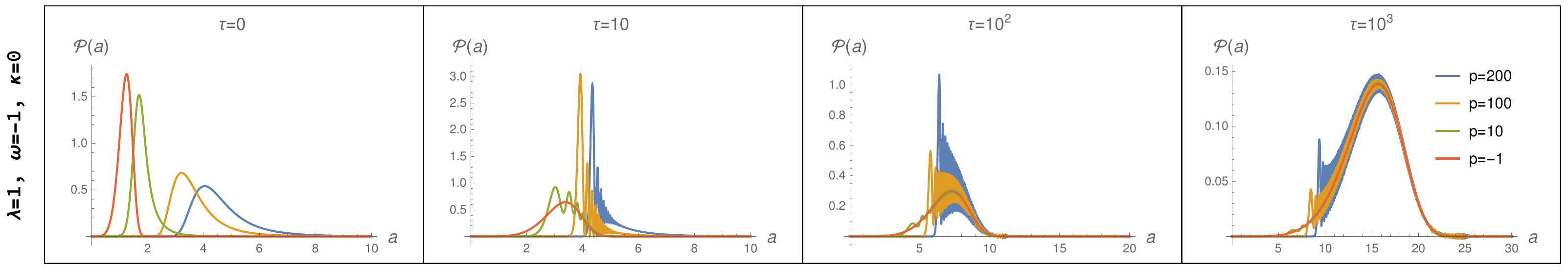}\vspace{-0.2cm}
	\includegraphics[width=\textwidth]{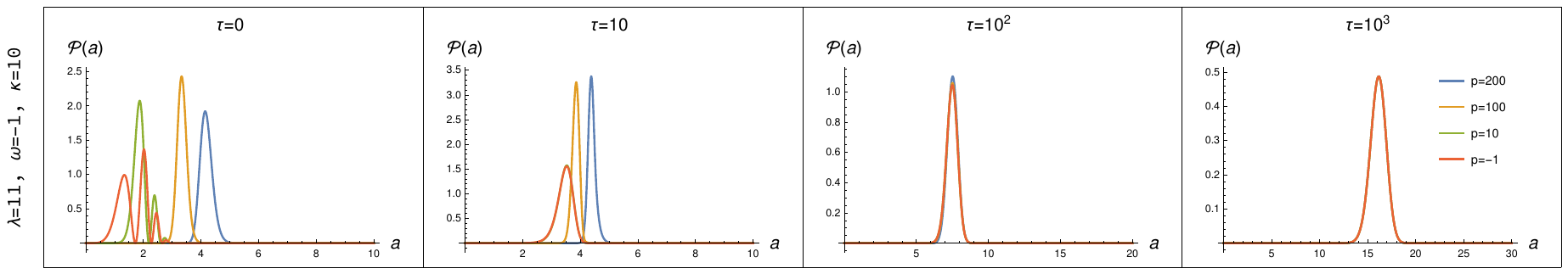}\vspace{-0.2cm}
	\includegraphics[width=\textwidth]{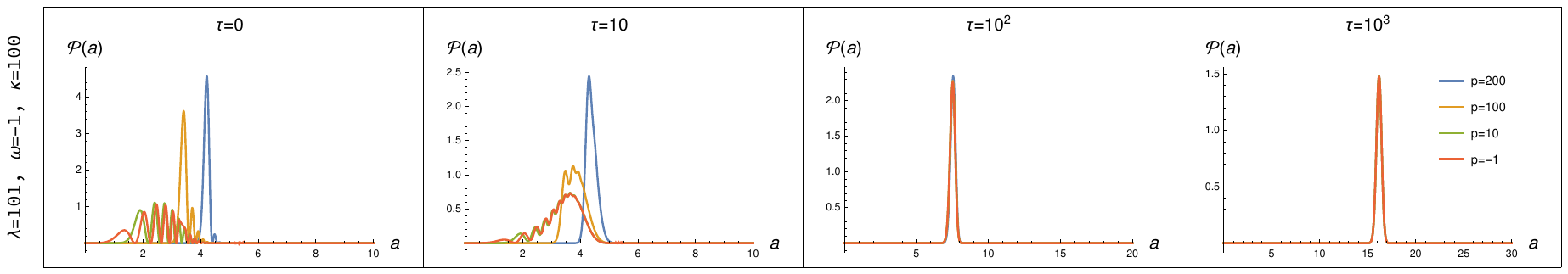}
	\caption{Time evolution of the probability distribution defined as $\mathcal{P}(a,\tau)=|\psi(a,\tau)|^2a^{1-3\omega-p-2q}$ associated with the wave packet in Eq. \eqref{WP} for different ordering parameters and the energy distribution of fixed mean energy and varying width in the case of a cosmological constant driven universe. Profiles of different colors represent different ordering choices. The parameter associated with the width of the distribution increases across the rows as the width decreases from top to bottom, and the time increases from left to right.}\label{Fig1}
\end{figure}

The behavior of the probability distribution associated with the wave packet, i.e., $\mathcal{P}(a,\tau)=|\psi|^2a^{1-3\omega-p-2q}$ is shown in Fig. \ref{Fig1}. Here, we have discussed the case of a cosmological constant driven universe, but the features observed here are generic in nature, and they appear for the other equation of state parameters as well, shown for the case of a dust dominated universe in \ref{App1}. We have plotted the probability distribution as a function of scale factor for different ordering choices, at different stages of the evolution and for distribution of fixed mean energy but different widths.\footnote{The probability distribution is symmetric in $\tau$ and therefore represents a symmetric bounce, as will be shown in Sec. \ref{Sec5}. Here, we will focus our attention on the expanding branch.}

In the first row, we have the parameter choice that corresponds to an energy distribution of a large width. At the bounce point $\tau=0$, the probability distribution for different orderings has distinct profiles that peak at different values of the scale factor, hinting at the high sensitivity of the size of the universe at the bounce on the ordering choice. The profile corresponding to $p=-1$ peaks at the minimum value of the scale factor, and the bounce size increases as $|1+p|$ increases. As the universe expands, the probability distribution disperses in general, and profiles corresponding to large ordering parameters acquire an oscillatory character with chirp-like features, and they tend to envelop the profile for the lowest value of $p=-1$ (the probability distribution is a function of $|1+p|$). Far away from bounce, different ordering profiles completely envelop the profile for $p=-1$, and the probability distribution now peaks at the same scale factor value. For the parameter values under consideration, it seems that the oscillatory nature of the probability distribution is the distinguishing characteristic that separates the small $p$ case from the large $p$, and a large ordering parameter leads to a higher frequency and higher amplitude of the oscillations.

However, the situation reverses as we increase $\kappa$. The profiles at the bounce start to attain oscillatory features, whereas the late-time profiles start to lose the oscillatory features. The profiles for different ordering are distinct again at the bounce point, and at a later time, they merge onto the same profile, leading to the understanding that the ordering ambiguity has no imprint at a late time, in this case. As we continue to increase the parameter $\kappa$, i.e., for sharper energy distribution, the oscillatory features at the bounce keep on enhancing, and the oscillatory behavior persists away from the bounce as well. 

From the analysis of the probability distribution, it is apparent that the ordering effects are most pronounced at the bounce point, where the probability distribution for different ordering parameters has different characteristics. During the later stages of the expansion, the imprints of ordering are apparent only for a broadly peaked energy distribution, manifested by the oscillatory nature of the probability distribution profiles, which persists away from the singularity. It is interesting to see whether the oscillatory behavior of the probability distribution at the late time can be captured by the expectation value of any observable. The behavior of probability distribution for varying mean energy and $-1<\kappa<0$ is discussed in \ref{App1}.

The evolution of the wave packet for a sharply peaked energy distribution and a small ordering parameter $(\kappa=100,\; p=-1)$ resembles the case of a Gaussian state for a free particle that is reflected from a hard boundary \cite{Andrews_WP,ROBINETT20041}. Initially, in the collapsing phase, i.e. $\tau\ll 0$, the probability distribution is a single-peaked profile. As the state evolves toward the bounce point, it starts attaining oscillatory features and is highly oscillatory at the bounce point. These oscillatory features disappear, giving a single peaked profile as the system evolves away from the bounce point. These oscillatory characteristics are conventionally attributed to the interference of the incoming part of the wave packet with the outgoing part \cite{Andrews_WP}. The same interpretation holds for the wave packet under consideration, giving us the familiar notion of quantum bounce \cite{Alexandre:2022ijm,Alexandre:2022npo,Gielen:2022dhg}, making a good case for the small value of the ordering parameter as the preferred choice. On the other hand, the behavior of the wave packet for a broadly peaked energy distribution is highly counterintuitive, and the origin and interpretation of the oscillatory features at late times are a mystery at this juncture. 

\section{Imprints of Hamiltonian ordering on the quantum dynamics}\label{Sec5}
\begin{figure*}
	\centering
	\begin{tabular}{|c|c|c|}
		\hline
		\includegraphics[width=0.3\textwidth]{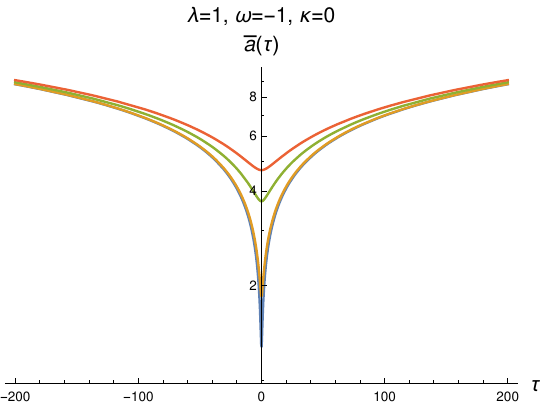} & \includegraphics[width=0.3\textwidth]{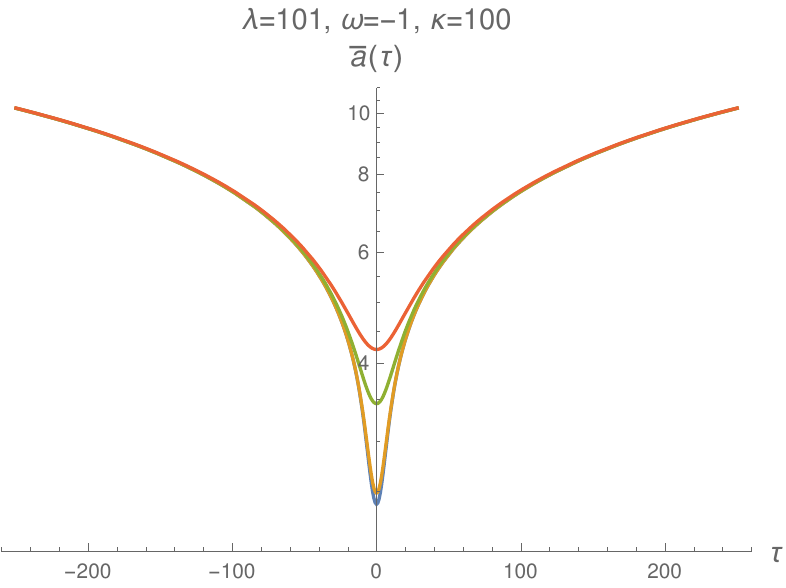}& \includegraphics[width=0.3\textwidth]{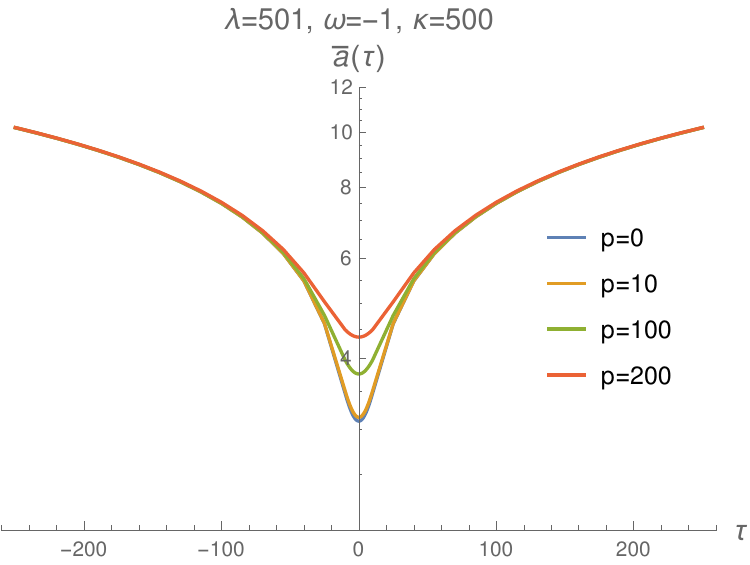}\\
		\hline
		\includegraphics[width=0.3\textwidth]{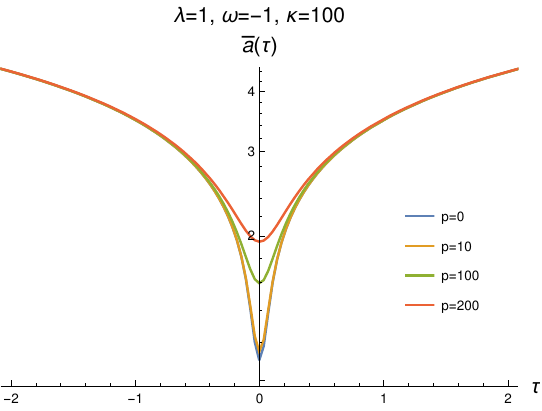} &
		\includegraphics[width=0.3\textwidth]{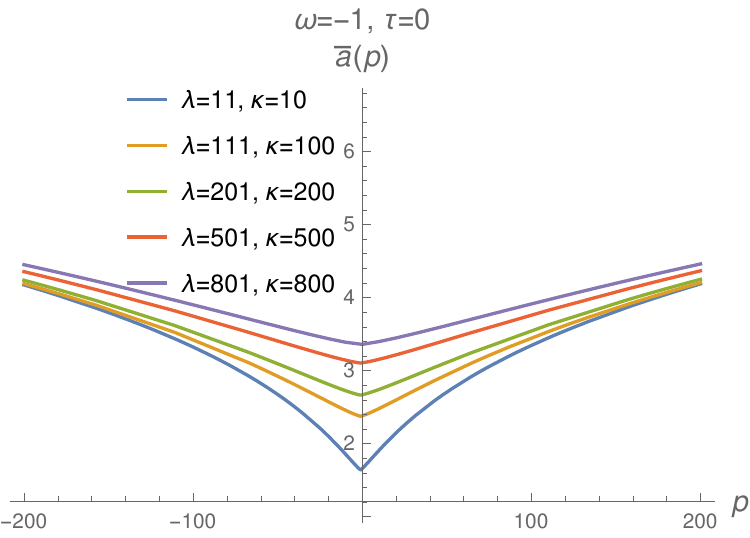} &  \includegraphics[width=0.3\textwidth]{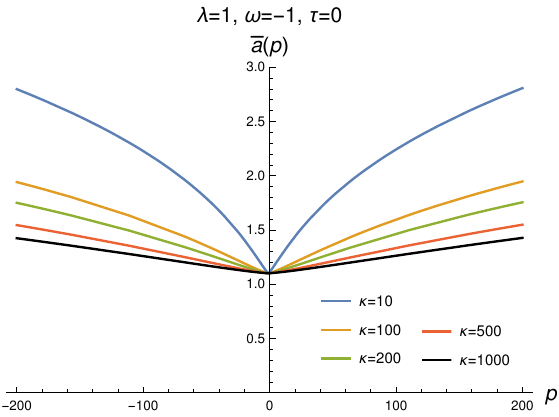} \\
		\hline
	\end{tabular}
	\caption{In the first row, the evolution of the expectation value of the scale factor for the cosmological constant driven universe with different widths of the energy distribution but fixed mean energy. Curves of different colors represent different orderings of the Hamiltonian constraint. In the first frame of the second row, we have the case for a different mean energy but the same width as for the middle frame of the first row. In the last two frames of the second row, we plot the expectation value of the scale factor at the bounce point as a function of the ordering parameter for different values of the shape parameter represented by the curves of different colors. In the middle frame of the second row, we have kept the mean energy fixed, and the width of the energy distribution decreases as $\lambda$ and $\kappa$ increase. In the last frame, we have fixed the parameter $\lambda$, so the mean energy also changes as we change $\kappa$.}\label{fig:sf}
\end{figure*}  
We are interested in the ordering dependence of the quantum dynamics of the universe, introduced by the expectation value of the scale factor. Due to the complicated nature of the wave packet, we resort to numerical computation of the expectation value using Mathematica, and the analytical behavior is discussed in \ref{App} for a subclass of wave packets with choice $\kappa=|1+p|/3(1-\omega)$. As alluded to in the discussion on the probability distribution, we are interested in the ordering dependence of the expectation values for two cases: a broadly peaked energy distribution, i.e., small $\kappa$, and a sharply peaked energy distribution for large $\kappa$. We consider the case of a cosmological constant driven universe, with the understanding that the generic features are present for other fluid choices as well, where the case of dust dominated universe is discussed in \ref{App2}. 

In Fig. \ref{fig:sf}, we have plotted the expectation value of the scale factor, and it follows the trend as anticipated from the discussion about the probability distribution. The system tunnels from a collapsing branch to an expanding branch and undergoes a symmetric quantum bounce. The signature of ordering ambiguity is most pronounced at the bounce point, where scale factor expectation has a non-zero minimum and profiles for different ordering merge together for large $|\tau|$. Furthermore, we see that the expectation value of the scale factor is insensitive to the oscillatory nature of the probability distribution for the widely peaked energy distribution and follows the classical behavior for large $\tau$ regardless of the ordering or width of the energy distribution. As we decrease the width of the energy distribution, keeping its mean energy fixed, we see that the profiles of different orderings begin to merge, and one can expect a single profile for all ordering choices in the limit $\kappa\rightarrow\infty$. The time window for which the ordering effects are relevant shrinks as we increase the mean energy by increasing the parameter $\kappa$ but fixing the parameter $\lambda$, as we see in the first frame of the second row of Fig. \ref{fig:sf}. For a subclass of wave packets considered in \cite{Sahota2023}, we analytically showed that the time scale for which the quantum effects are relevant is, in fact, related to the mean energy.

However, from the discussion thus far, it is not clear whether the ordering imprints completely disappear from the dynamics of the universe for a sharply peaked energy distribution. The signatures of ordering ambiguity are most pronounced at the bounce point, and therefore, it is prudent that we investigate the sensitivity of the expectation value of the scale factor at the bounce on the ordering of the Hamiltonian. We plot the expectation value of the scale factor at the bounce as a function of the ordering parameter for different values of the shape parameters for the fixed energy in the middle frame of the second row in Fig. \ref{fig:sf}, while the mean energy also changes in the last frame according to \eqref{disP} as we keep $\lambda$ fixed.

In both cases, the bounce size is minimum for $p=-1$ and is a monotonically increasing function of $|p+1|$. As we increase the sharpness of the energy distribution, this minimum bounce size continues to increase, and the slope of dependence on the ordering decreases. We can anticipate that the dependence on the ordering will be minimal in the limit of $\kappa\rightarrow\infty$. However, it is not apparent from the middle frame of the second row in Fig. \ref{fig:sf} that the ordering imprints indeed wash away for the fixed mean energy. On the other hand, for fixed $\lambda$, as we continue to increase $\kappa$, $\Bar{a}_0(p)$ curve continues to flatten, and we can anticipate a flat curve parallel to the p-axis for a sharply peaked state in this case. Surprisingly, the minimum value of the bounce size for the parameter $p=-1$ remains the same for different values of the shape parameters and for fixed $\lambda$. All other ordering choices approach this value in the limiting case of $\kappa\rightarrow\infty$. Since the mean energy changes as we change $\kappa$, for a sharply peaked distribution with a large mean energy, the ordering imprints are minimal.

\begin{figure}[!t]
	\centering
	\begin{tabular}{|c|c|}
		\hline
		\includegraphics[width=0.48\textwidth]{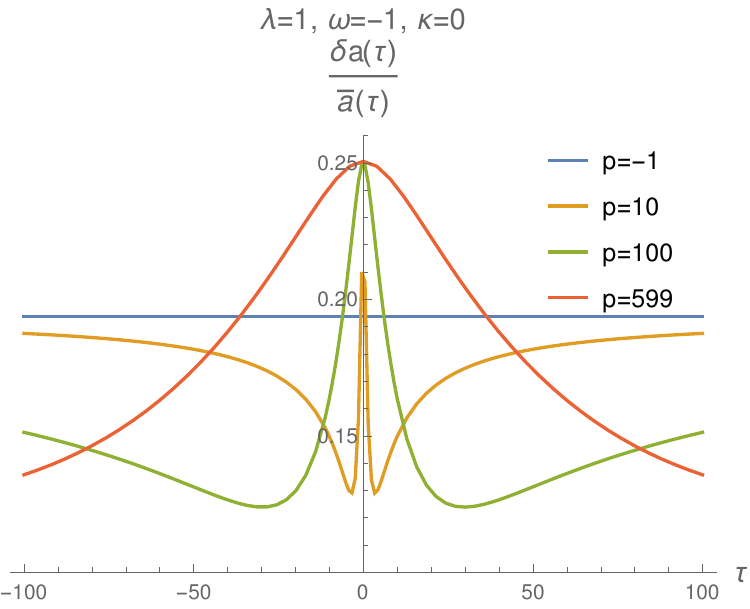} & 
            \includegraphics[width=0.48\textwidth]{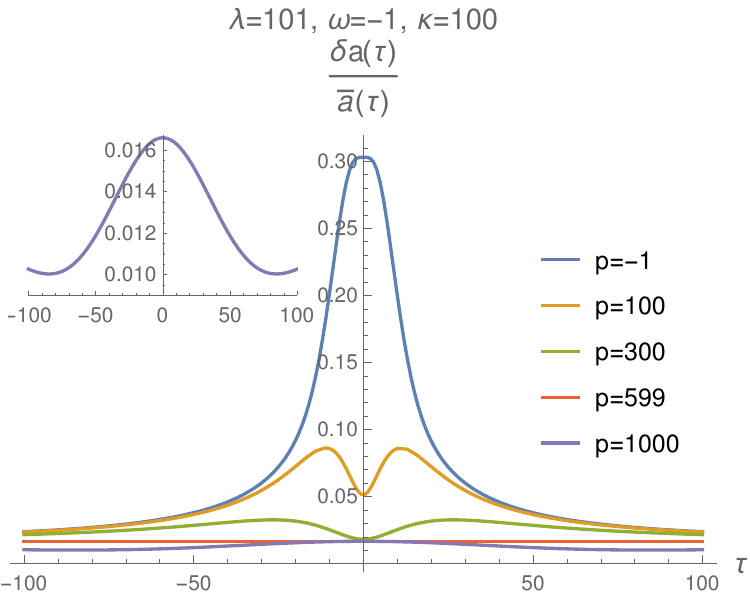} \\
		\hline
	\end{tabular}
	\caption{Relative standard deviation in scale factor as a function of time for different ordering choices, for a cosmological constant driven universe. In the first frame, we have the case of a broadly peaked distribution, and a sharply peaked distribution in the second frame. }\label{fig:dsf}
\end{figure}

Next, we are interested in the dynamics of fluctuations in the scale factor for different ordering choices. We plot the relative standard deviation in the scale factor as a function of time for the broadly peaked energy distribution in the first frame of Fig. \ref{fig:dsf} and the sharply peaked distribution in the second frame. The relative standard deviation turns out to be independent of time for the parameter choice $\kappa=|1+p|/3(1-\omega)$,\footnote{The states considered in \cite{Sahota2023} corresponds to this choice.} and for other ordering choices, the relative standard deviation asymptote to this value far away from the bounce. The generic behavior of quantum fluctuations for a broadly peaked distribution is that it has a global maximum at the bounce that is sandwiched between two global minima, and it asymptotically reaches the limiting value at late time\footnote{This observation is clearer for a dust dominated universe, see middle and last frames of second row in Fig. \ref{fig:sfB} in the \ref{App2}, as the numerical integration for large $\tau$ is unstable due to the oscillatory nature of the integrand in the case of a cosmological constant driven universe.}. For a larger ordering parameter, the location of global minima is pushed away from the bounce, although the magnitude of global maximum does not change. A point of reflection, in this case, is that the signatures of ordering ambiguity persist far away from the bounce, where the quantum fluctuations are not decaying but are increasing toward the limiting value asymptotically. Another observation is that the magnitude of the quantum fluctuations near the bounce is of the same order as compared to the late-time quantum fluctuations for all ordering choices.

The quantum fluctuations for sharply peaked states show a further intriguing character. For the shape parameter $\kappa=100$, the limiting case of constant quantum fluctuation is for the ordering choice $p=599$. The other ordering choices again asymptote to this value far away from the bounce. The nature of the extremum of quantum fluctuations at the bounce depends on the ordering choice. For the ordering parameter $p=-1$, the quantum fluctuations have a maximum at the bounce and monotonically decrease for large $|\tau|$. For $|1+p|<599$, the quantum fluctuations have a local minimum at the bounce sandwiched between two global maxima, where the magnitude of global minima is decreasing as $|1+p|$ increases, reaching the limiting value for $p\sim300$. For the limiting case $p=599$, the relative quantum fluctuations are constant, and as $p$ increases further, the quantum fluctuations at bounce attain a global maximum that is sandwiched between the global minima, as shown in the inset of the second frame in Fig. \ref{fig:dsf}. In this case, the asymptotic behavior of the quantum fluctuation far away from the bounce is a distinguishing feature, where it decreases towards the limiting value for $|1+p|<599$, and it increases towards the limiting value for $|1+p|>599$. Furthermore, for certain ordering choices, i.e., small $p$, quantum fluctuations near the bounce are substantially larger than late-time quantum fluctuations. On the other hand, for large $p$ the magnitude of quantum fluctuations near the bounce is of the same order as compared to the late-time quantum fluctuations. In general, the magnitude of late-time quantum fluctuations decreases as we decrease the width of the energy distribution.

On a broader note, the notion of singularity resolution is closely linked to the choice of clock in quantum theory, as conjectured by Gotay and Demaret \cite{Gotay_Singularity}. It states that the slow clock choices, such as the perfect fluid clock, that encounter the singularity in finite time are expected to resolve singularity, whereas the fast clocks, such as volume clock or scalar field clock, are expected to encounter singularity in quantum theory \cite{Malkiewicz:2019azw,Gielen:2020abd,Gielen:2021igw}. The quantum bounce observed in this model is an artifact of the boundary conditions of the quantum theory, where the unitary dynamics in the fluid variable lead to the tunneling from the collapsing branch to the expanding branch. In this analysis, we work toward a better understanding of the bouncing geometry in this framework by checking its robustness against the operator ordering ambiguity of canonical quantum gravity. We further investigated whether the expectation value of the scale factor provides an appropriate quantum-corrected trajectory by checking the behavior of relative fluctuations in the scale factor. Our results are encouraging as the notion of quantum-corrected spacetime introduced through the expectation value of scale factor is for the sharply peaked states.

At this point, a comment on the ordering scheme used in the literature is in order. The energy distribution considered in this work is used in numerous related works \cite{Alvarenga:1998wx,Alvarenga:2001nm,Batista:2001ti,hajicek_singularity_2001,Pedram:2007iv,Pedram:2007ck,Pedram:2007ud,Pedram:2007fm}. To obtain analytical results, one needs to make a choice about the distribution parameter and the ordering parameter similar to the one followed in \cite{kiefer_singularity_2019,Sahota_Infrared,Sahota2023}, where the order of the Bessel function in Eq. \eqref{Stast} is chosen to be equal to the parameter $\kappa$. The issue is that once we choose an ordering scheme that generically corresponds to small $p$, the width of the distribution is large by default. Therefore, the ordering ambiguity will be relevant in addition to large quantum fluctuations, and the notion of the quantum-corrected spacetime is, therefore, ill-defined in these analyses. 

In conclusion, near-bounce quantum dynamics is highly sensitive to the ordering chosen, and its imprint is most pronounced for states constructed from a broadly peaked energy distribution. The signature of operator ordering ambiguity is minimal, and the quantum fluctuations are small for states sharply peaked on a quantum-corrected trajectory. These states are of particular importance, as they are at the center of the dressed-metric approach in \cite{Ashtekar:2009mb}, and the notion of the quantum-corrected spacetime is well-defined for these states \cite{Sahota2023}. Thus, we have shown that the ordering effects left a minimal imprint in the case of the class of states relevant to the semiclassical analysis. On the other hand, we expect the ordering to be irrelevant for a Dirac delta-like state, i.e., with $\Delta E=0$ or the stationary state in \eqref{Stast}. However, these states are not part of the Hilbert space but the space of distributions, i.e., the antidual space in the Gelfand triplet \cite{Madrid_2005}, in the same spirit as plane waves. The realistic scenario is that we have a state with finite although small $\Delta E/\bar{E}$, and therefore, the system will always have some sensitivity to the ordering chosen.

\section{Discussion}\label{Sec6}
We have investigated imprints of ordering ambiguity on the dynamics of a perfect fluid dominated quantum universe. A general ordering scheme is employed for the Hamiltonian operator and a wave packet is constructed that represents the quantum bounce. We start with the analysis of the probability distribution associated with the wave packet. The wave packet with sharply peaked energy distribution and small ordering parameter has a direct correspondence with a Gaussian state reflecting from a hard wall; in other cases, the behavior of the probability distribution is highly non-trivial. In the case of wave packets with a broadly peaked energy distribution, the probability distribution has oscillatory features at late times for a large ordering parameter, providing a possible avenue to investigate the signatures of ordering ambiguity.

The expectation value of the scale factor represents a robust, symmetric bounce with appropriate classical behavior away from the singularity. The quantum dynamics of the universe turn out to be insensitive to the oscillatory character of the probability distribution at a late time in the expanding phase. As the width of the energy distribution decreases, the profiles with different orderings tend to merge, and we expect the ordering imprints to wash away in this regime. The ordering imprints are most pronounced at the bounce, and we investigate the dependence of the bounce size on the ordering parameter for different values of the shape parameter. We find that the scale factor expectation is insensitive to the ordering chosen for sharply peaked states. The analysis suggests that the ordering imprints are washed away in the limit of the energy distribution of the vanishing width and the large mean energy. Moreover, we show that the quantum fluctuations in scale factor are of the same order throughout the evolution of the universe for broadly peaked states, whereas they are decaying to small albeit finite values far away from bounce for sharply peaked states. However, the ordering choice does dictate the near-bounce behavior of quantum fluctuations, even for sharply peaked states.

We show that quantum ambiguities are of little relevance and that late-time quantum fluctuations are small for a universe with a well-defined energy, i.e., $\Delta E\ll\overline{E}$. The conjecture of sharply peaked state turns out to be a savior from the quantization ambiguities in the effective geometry approach \cite{Ashtekar:2009mb,Sahota2023} in the case of this toy model, and the expectation value of scale factor provides a consistent quantum-corrected spacetime. However, the ordering choice will leave some signature on the cosmological observables through their imprint on the size from which the universe bounces off and the quantum fluctuations in the scale factor for physical states (which are not infinitely peaked with $\Delta E=0$). In standard quantum cosmology analysis, the Vilenkin or Laplace-Beltrami ordering choice leads to large quantum fluctuations and leaves an imprint of ordering ambiguity for states considered in \cite{Alvarenga:1998wx,Alvarenga:2001nm,Batista:2001ti,Pedram:2007iv,Pedram:2007ck,Pedram:2007ud,Pedram:2007fm}. On a side note, one wonders whether the conjecture of sharply peaked states can save other canonical approaches to quantum gravity, e.g., loop quantum cosmology \cite{Ashtekar:2009mb,Agullo:2012fc,Agullo:2012sh}, from the operator ordering ambiguities as well. 

\section{Acknowledgments}
    HSS acknowledges the financial support from the University Grants Commission, Government of India, in the form of Junior Research Fellowship (UGC-CSIR JRF/Dec-2016 /503905). HSS thanks Kinjalk Lochan and Vikramaditya Mondal for the careful reading of manuscript, and for their comments and suggestions. HSS is grateful to Patrick Peter and Przemyslaw Ma{\l}kiewicz for the helpful discussions during the early stage of this work and their hospitality during his visit to IAP Paris and NCBJ Warsaw. HSS is grateful to the organizers of the conference `Time and Clocks' for the hospitality at Physikzentrum Bad Honnef and thanks Martin Bojowald for his comments on my previous work that initiated this project.

\appendix
    % With this wave packet, the time derivative of the scale factor expectation takes the form, 
    % \begin{align}
    %     \frac{d\overline{a}(\tau)}{d\tau}&=\int_0^\infty da\;a^{2-3\omega-p-2q}\frac{d|\psi(a,\tau)|^2}{d\tau}\nonumber\\
    %     &\propto\tau\left(\frac{\lambda ^2}{4}+\tau ^2\right)^{\frac{| p+1| }{3 (\omega -1)}-3}\int_0^\infty da\;a^{2-3\omega-p-2q}  a^{1+p+2q+|1+p|-3 \omega }  e^{-\frac{2 \lambda  a^{3(1- \omega )}}{9 (1-\omega)^2 \left(\frac{\lambda ^2}{4}+\tau ^2\right)}}\nonumber\\
    %     &\qquad\qquad\left(8 a^3 \lambda -3 (1-\omega ) a^{3 \omega } \left(\lambda ^2+4 \tau ^2\right) (| p+1| +3(1- \omega))\right)
    % \end{align}
    % which vanishes at the $\tau=0$, implying it is an extremum. Next, we check whether the extrema is a minimum or a maximum by looking at the behavior of the second derivative
    % \begin{align}
    %     \frac{d^2\overline{a}(\tau)}{d\tau^2}\bigg|_{\tau\rightarrow0}&=\int_0^\infty da\;a^{2-3\omega-p-2q}\frac{d^2|\psi(a,\tau)|^2}{d\tau^2}\bigg|_{\tau\rightarrow0}\nonumber\\
    %     &\propto\int_0^\infty da\;a^{2-3\omega-p-2q}a^{-3 \omega } e^{-\frac{8 a^{3-3 \omega }}{9 \lambda  (1-\omega)^2}} \left(8 a^3-3 \lambda  (1-\omega ) a^{3 \omega } (| p+1| +3 (1-\omega ))\right)\nonumber
    % \end{align}

\section{Evolution of wave packet}\label{App1}
\begin{figure}[H]
		\includegraphics[width=\textwidth]{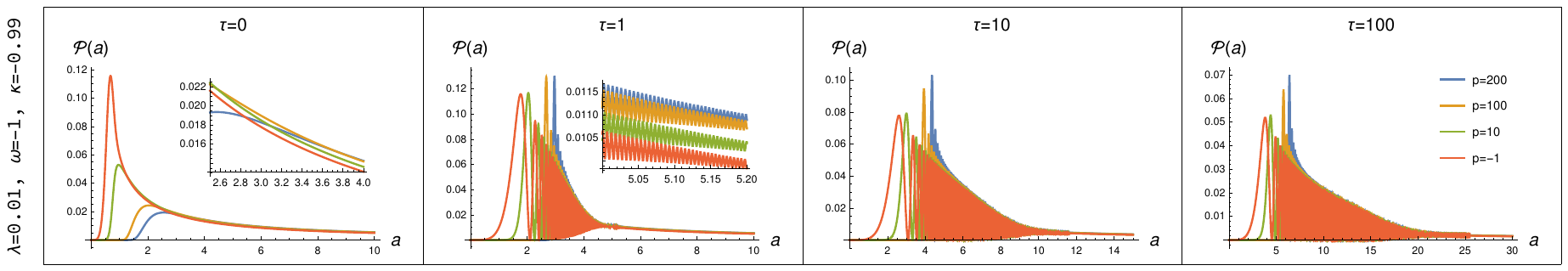}\vspace{-0.2cm}
		\includegraphics[width=\textwidth]{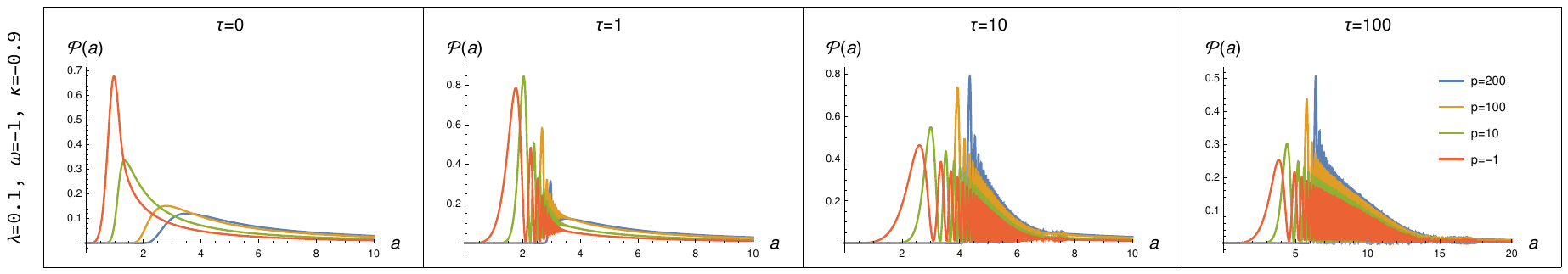}\vspace{-0.2cm}
	\caption{Time evolution of the probability distribution associated with the wave packet in Eq. \eqref{WP} for different ordering parameters and the energy distribution of fixed mean energy and broadly peaked width in the case of a cosmological constant driven universe. Profiles of different colors represent different ordering choices.}\label{Figapp}
\end{figure}

In this appendix, we discuss the evolution of wave packet for the cases not considered in the main text. In Fig. \ref{Figapp}, we have plotted the probability distribution for the case where $-1<\kappa<0$, where the width of the energy distribution is comparatively larger. In this case, we see that the behavior of probability distribution is even more peculiar with no apparent correspondence between different ordering choices at any stage of the evolution of the universe. 

\begin{figure}[h!]
	\includegraphics[width=\textwidth]{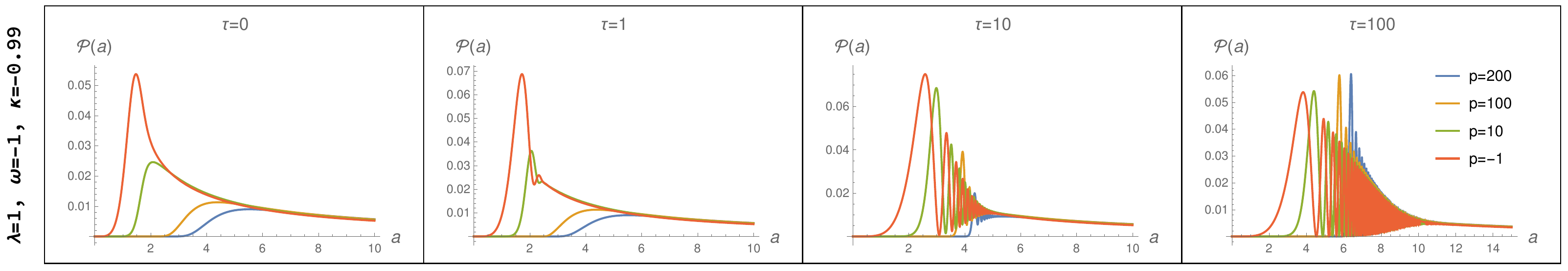}\vspace{-0.2cm}
	\includegraphics[width=\textwidth]{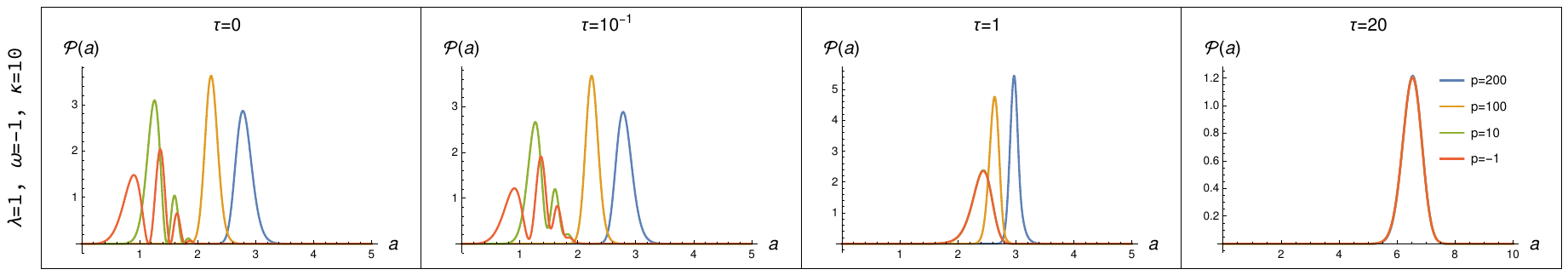}\vspace{-0.2cm}
	\includegraphics[width=\textwidth]{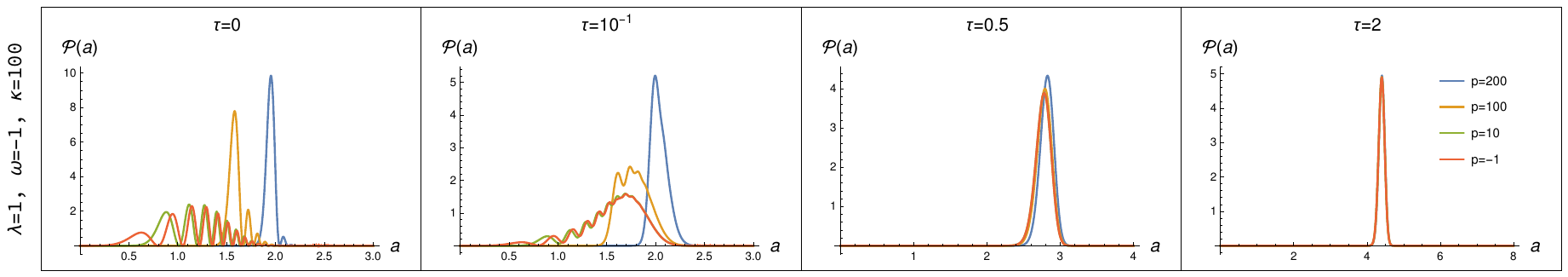}
	\caption{Time evolution of the probability distribution for different ordering parameters and the energy distribution of varying mean energy and varying width in the case of a cosmological constant driven universe.}\label{Fig2}
\end{figure}

\begin{figure}[h!]
	\includegraphics[width=\textwidth]{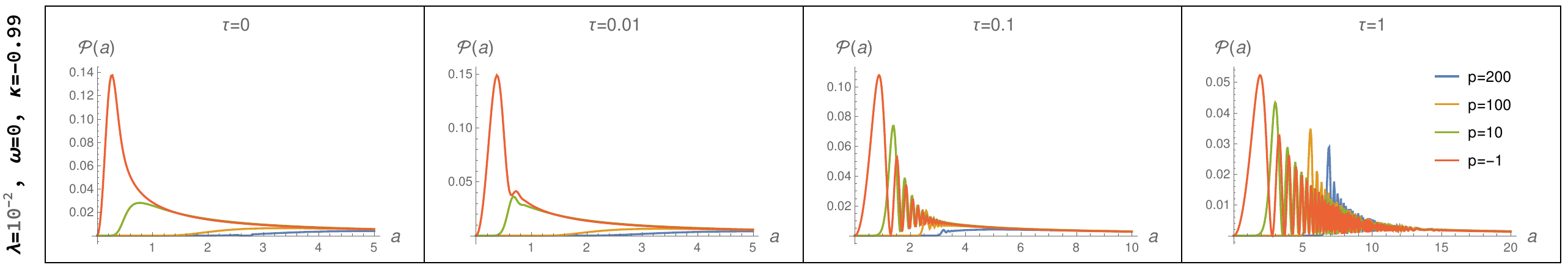}\vspace{-0.2cm}
	\includegraphics[width=\textwidth]{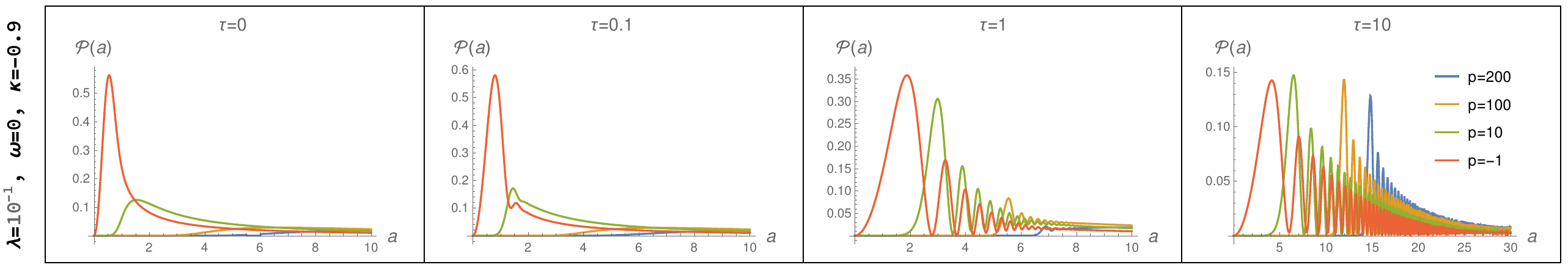}\vspace{-0.2cm}
	\includegraphics[width=\textwidth]{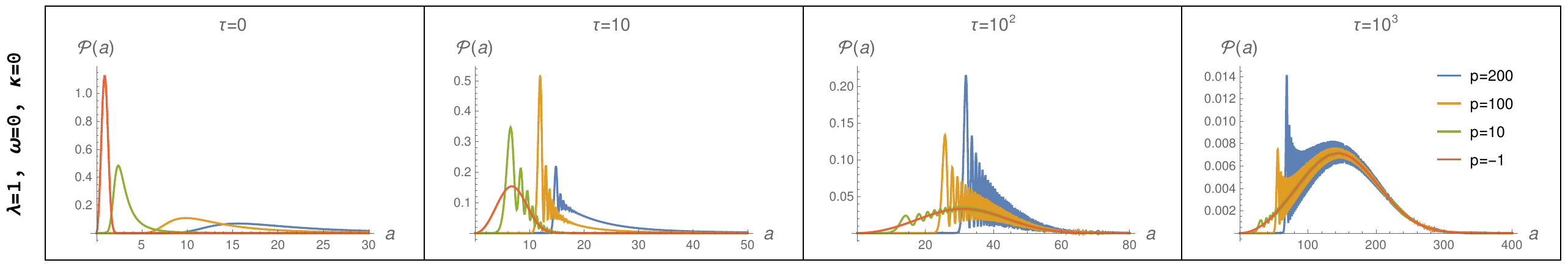}\vspace{-0.2cm}
	\includegraphics[width=\textwidth]{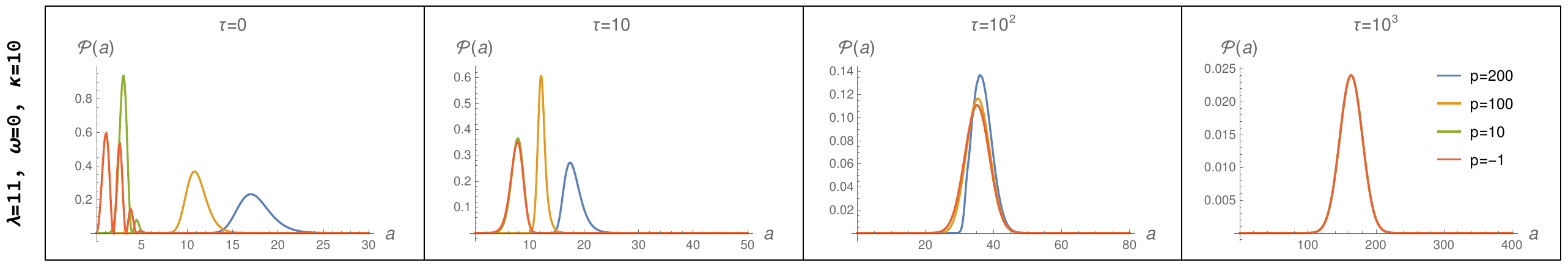}\vspace{-0.2cm}
	\includegraphics[width=\textwidth]{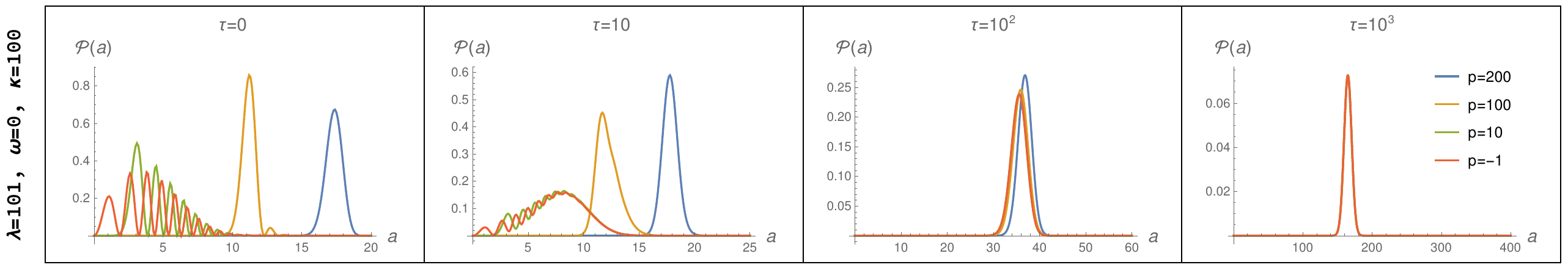}
	\caption{Time evolution of the probability distribution for different ordering parameters and the energy distribution of fixed energy and of varying width in case of a matter dominated universe.}\label{Fig3}
\end{figure}

In Fig. \ref{Fig2}, we change the mean energy of the distribution as well as its width. We see that the dynamics of the probability distribution have the same generic features as we have seen for the previous case, although notably, the ordering imprints for a sharply peaked state sustain for a shorter duration only. Again, we have late-time oscillatory features in the probability distribution for the broadly peaked energy distribution, and for the sharply peaked energy distribution, we have oscillatory features at the bounce. Therefore, it seems that the behavior of the probability distribution is sensitive to the sharpness of the energy distribution rather than to the mean energy. In Fig. \ref{Fig3}, we show that the characteristic features in the dynamics of the probability distribution are present for a dust dominated universe as well. 

\section{Consistency of quantum bounce}\label{App}
In this Appendix, we analytically show that the expectation value of the scale factor represents a bouncing universe with appropriate behavior far away from the bounce. Ideally, one need to show $\displaystyle \frac{d\overline{a}(\tau)}{d\tau}\bigg|_{\tau\rightarrow0}=0$, $\displaystyle \frac{d^2\overline{a}(\tau)}{d\tau^2}\bigg|_{\tau\rightarrow0}>0$, and $\overline{a}(\tau\rightarrow0)\neq 0$. For the given state, the expectation value of the scale factor is
    \begin{align}
        \overline{a}(\tau)=\int_0^\infty da\;a^{1-3\omega-p-2q}\psi^*(a,\tau)a\psi(a,\tau)
    \end{align}
    For the wave packet under consideration \eqref{WP}, it is harder to show that these conditions hold owing to its complicated form. However, this can be done for the case where we make the choice $\kappa=|1+p|/3(1-\omega)$ leading to the wave packet in the form of a generalized Gaussian
    \begin{align}
			\psi(a,\tau)=&\sqrt{\frac{3(1-\omega)}{\Gamma\left(\frac{| p+1| }{3 (1-\omega )}+1\right)}}\left(\frac{\frac{\sqrt{2\lambda}}{3 (1-\omega )}}{\left(\frac{\lambda }{2}-i \tau \right)}\right)^{\frac{| p+1| }{3 (1-\omega )}+1}a^{\frac{1}{2} (| p+1| +p+2 q+1)}\, e^{-\frac{2 a^{3 (1-\omega )}}{9 (1-\omega )^2 \left(\frac{\lambda }{2}-i \tau \right)}}.
    \end{align}
    The expectation value of the scale factor can be computed analytically, leading to the expression
    \begin{align}
        \overline{a}(\tau)=\left(\frac{9 (1-\omega )^2 \left(\lambda ^2+4 \tau ^2\right)}{8 \lambda }\right)^{\frac{1}{3 (1-\omega )}}\frac{\Gamma \left(\frac{4-3 \omega +|1+ p|}{3 (1-\omega)}\right)}{\Gamma \left(1+\frac{| 1+p| }{3(1-\omega)}\right)}.
    \end{align}
    Asymptotically, the scale factor expectation far away from the classical singularity behaves classically
    \begin{align}
        \overline{a}(\tau)\big|_{\tau^2\gg\lambda^2}=\left(\frac{9(1-\omega)^2}{2\lambda}\right)^{\frac{1}{3(1-\omega)}}\frac{\Gamma \left(\frac{4-3 \omega +|1+ p|}{3 (1-\omega)}\right)}{\Gamma \left(1+\frac{| 1+p| }{3(1-\omega)}\right)}\tau^{\frac{2}{3(1-\omega)}}=\braket{a_{cl}(\tau)}
    \end{align}
     where the object in the last equality is the ensemble average of the classical scale factor in \eqref{sfC} with distribution \eqref{dis}. The expectation value of the scale factor has a global minimum at $\tau=0$, as can be seen from the derivative test
     \begin{align}
         \frac{d\overline{a}(\tau)}{d\tau}=\frac{\tau \left((1-\omega )^2/\lambda\right)^{\frac{1}{3-3 \omega }}}{(\lambda ^2+4 \tau ^2)^{1-\frac{1}{3-3 \omega }}}\frac{2^{3-\frac{1}{1-\omega }} 9^{\frac{1}{3 (1-\omega )}} \Gamma \left(\frac{-3 \omega +| p+1| +4}{3 (1-\omega )}\right)}{| p+1|  \Gamma \left(\frac{| p+1| }{3 (1-\omega )}\right)}\xrightarrow[]{\tau=0}0,
     \end{align}
     \begin{align}
         \frac{d^2\overline{a}(\tau)}{d\tau^2}\bigg|_{\tau\rightarrow0}=\frac{2^{\frac{1}{\omega -1}+3} 3^{\frac{2}{3-3 \omega }-1} \left(\lambda  (1-\omega )^2\right)^{\frac{1}{3-3 \omega }} \Gamma \left(\frac{-3 \omega +| p+1| +4}{3-3 \omega }\right)}{\lambda ^2 (1-\omega ) \Gamma \left(\frac{| p+1| }{3-3 \omega }+1\right)}>0.
     \end{align}

\section{Evolution of the dust dominated quantum universe}\label{App2}
\begin{figure}[H]
	\centering
	\begin{tabular}{|c|c|c|}
		\hline
		\includegraphics[width=0.3\textwidth]{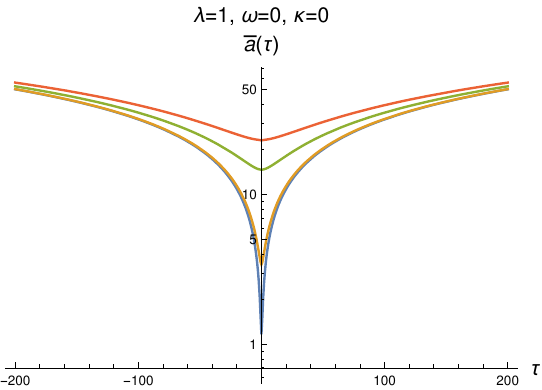} & \includegraphics[width=0.3\textwidth]{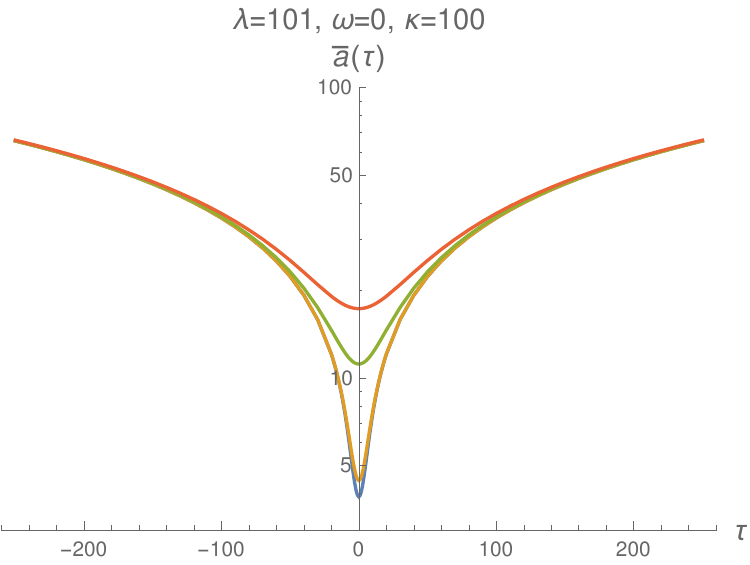} &
		\includegraphics[width=0.3\textwidth]{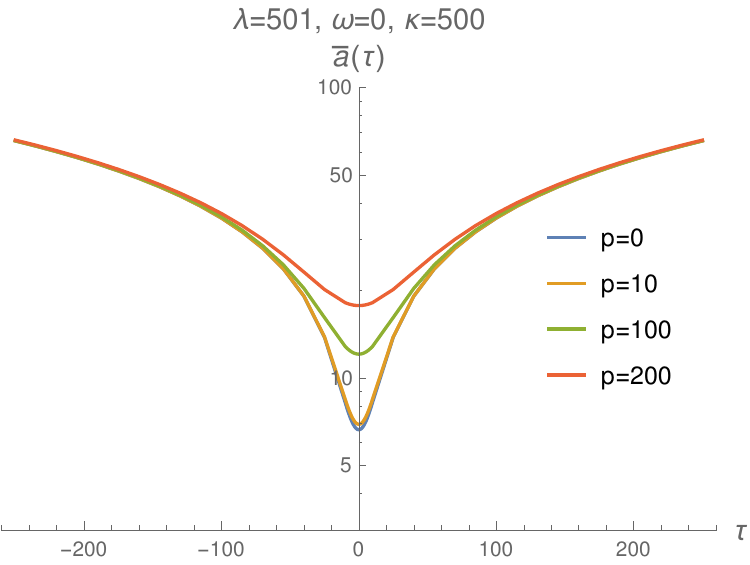} \\
		\hline
		\includegraphics[width=0.3\textwidth]{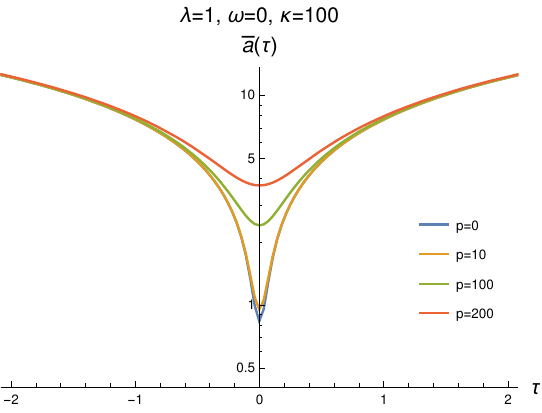} &\includegraphics[width=0.3\textwidth]{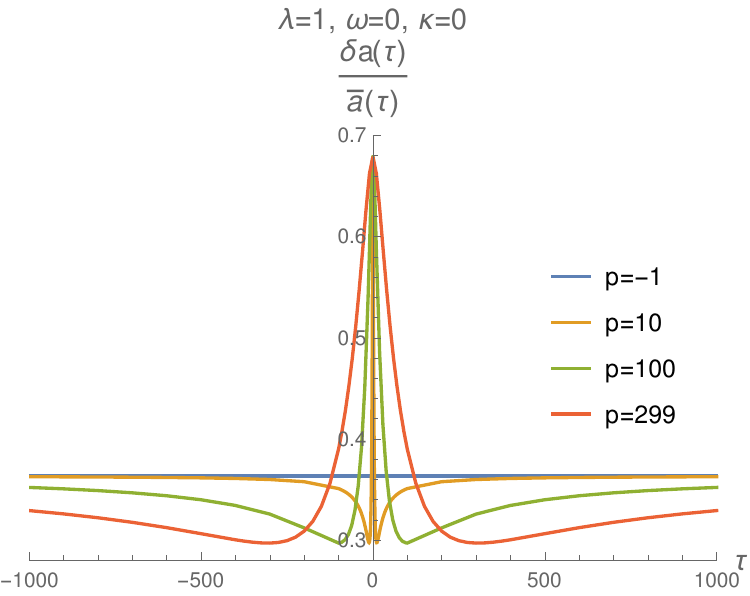} &  \includegraphics[width=0.3\textwidth]{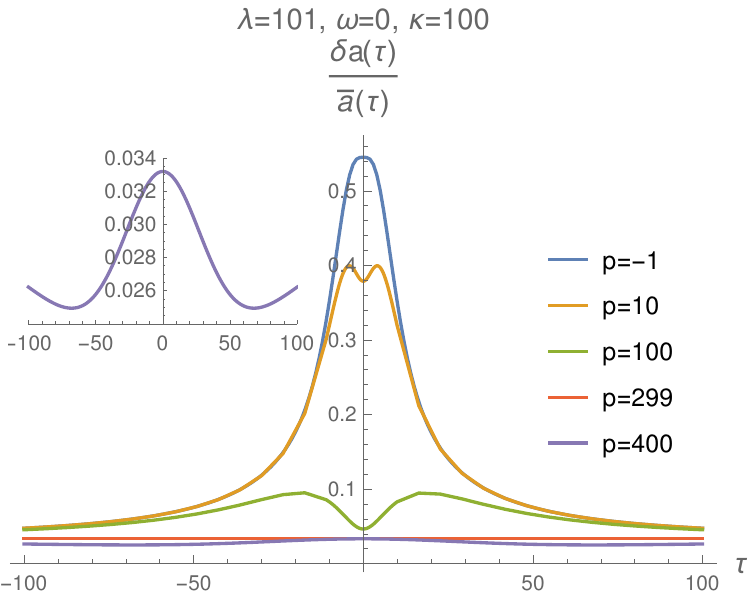}  \\
		\hline
	\end{tabular}
	\caption{The expectation value of the scale factor for the cosmological constant driven universe in the first row and the dust dominated universe in the second row with different values of the shape parameter $\kappa=0$, $\kappa=100$ and $\kappa=500$, but fixed mean energy. Curves with different colors represent different orderings of the Hamiltonian constraint. In the third row, we have the case for different mean energy.}\label{fig:sfB}
\end{figure}

In this appendix, we discuss the quantum evolution of a dust dominated universe. In the first row of Fig. \ref{fig:sfB}, we have plotted the expectation value of the scale factor for different ordering choices and for the energy distribution of fixed mean energy and decreasing width. We see the results follow the cosmological constant driven universe considered in the main text, the ordering imprints are most pronounced at the bounce, the domain of ordering dependence is shrinking, and the different ordering profiles tend to merge with decreasing width. Moreover, as we increase the mean energy, the time window of ordering dependence decreases by several orders of magnitude. In the second row of Fig. \ref{fig:sfB}, we plot the quantum fluctuations in the scale factor in the last two frames. In this case, the numerical computation for broadly peaked state and far away from bounce is stable, and we see that the fluctuations for different ordering choices tend to limiting case. For quantum fluctuations as well, the dust-dominated universe has the same characteristic features as in the case of the cosmological constant driven universe.


\begin{thebibliography}{37}%
	\makeatletter
	\providecommand \@ifxundefined [1]{%
		\@ifx{#1\undefined}
	}%
	\providecommand \@ifnum [1]{%
		\ifnum #1\expandafter \@firstoftwo
		\else \expandafter \@secondoftwo
		\fi
	}%
	\providecommand \@ifx [1]{%
		\ifx #1\expandafter \@firstoftwo
		\else \expandafter \@secondoftwo
		\fi
	}%
	\providecommand \natexlab [1]{#1}%
	\providecommand \enquote  [1]{``#1''}%
	\providecommand \bibnamefont  [1]{#1}%
	\providecommand \bibfnamefont [1]{#1}%
	\providecommand \citenamefont [1]{#1}%
	\providecommand \href@noop [0]{\@secondoftwo}%
	\providecommand \href [0]{\begingroup \@sanitize@url \@href}%
	\providecommand \@href[1]{\@@startlink{#1}\@@href}%
	\providecommand \@@href[1]{\endgroup#1\@@endlink}%
	\providecommand \@sanitize@url [0]{\catcode `\\12\catcode `\$12\catcode
		`\&12\catcode `\#12\catcode `\^12\catcode `\_12\catcode `\%12\relax}%
	\providecommand \@@startlink[1]{}%
	\providecommand \@@endlink[0]{}%
	\providecommand \url  [0]{\begingroup\@sanitize@url \@url }%
	\providecommand \@url [1]{\endgroup\@href {#1}{\urlprefix }}%
	\providecommand \urlprefix  [0]{URL }%
	\providecommand \Eprint [0]{\href }%
	\providecommand \doibase [0]{https://doi.org/}%
	\providecommand \selectlanguage [0]{\@gobble}%
	\providecommand \bibinfo  [0]{\@secondoftwo}%
	\providecommand \bibfield  [0]{\@secondoftwo}%
	\providecommand \translation [1]{[#1]}%
	\providecommand \BibitemOpen [0]{}%
	\providecommand \bibitemStop [0]{}%
	\providecommand \bibitemNoStop [0]{.\EOS\space}%
	\providecommand \EOS [0]{\spacefactor3000\relax}%
	\providecommand \BibitemShut  [1]{\csname bibitem#1\endcsname}%
	\let\auto@bib@innerbib\@empty
	%</preamble>
	\bibitem [{\citenamefont {Kiefer}(2012)}]{kiefer_quantum_2012}%
	\BibitemOpen
	\bibfield  {author} {\bibinfo {author} {\bibfnamefont {C.}~\bibnamefont
			{Kiefer}},\ }\href@noop {} {\emph {\bibinfo {title} {Quantum gravity}}},\
	\bibinfo {edition} {third edition}\ ed.,\ \bibinfo {series} {International
		series of monographs on physics}\ No.\ \bibinfo {number} {155}\ (\bibinfo
	{publisher} {Oxford University Press},\ \bibinfo {address} {Oxford},\
	\bibinfo {year} {2012})\BibitemShut {NoStop}%
	\bibitem [{\citenamefont {DeWitt}(1967)}]{dewitt_quantum_1967}%
	\BibitemOpen
	\bibfield  {author} {\bibinfo {author} {\bibfnamefont {B.~S.}\ \bibnamefont
			{DeWitt}},\ }\bibfield  {title} {\bibinfo {title} {Quantum {Theory} of
			{Gravity}. {I}. {The} {Canonical} {Theory}},\ }\href
	{https://doi.org/10.1103/PhysRev.160.1113} {\bibfield  {journal} {\bibinfo
			{journal} {Physical Review}\ }\textbf {\bibinfo {volume} {160}},\ \bibinfo
		{pages} {1113} (\bibinfo {year} {1967})}\BibitemShut {NoStop}%
	\bibitem [{\citenamefont {Tsamis}\ and\ \citenamefont
		{Woodard}(1987)}]{Tsamis_ordering}%
	\BibitemOpen
	\bibfield  {author} {\bibinfo {author} {\bibfnamefont {N.~C.}\ \bibnamefont
			{Tsamis}}\ and\ \bibinfo {author} {\bibfnamefont {R.~P.}\ \bibnamefont
			{Woodard}},\ }\bibfield  {title} {\bibinfo {title} {The factor-ordering
			problem must be regulated},\ }\href
	{https://doi.org/10.1103/PhysRevD.36.3641} {\bibfield  {journal} {\bibinfo
			{journal} {Phys. Rev. D}\ }\textbf {\bibinfo {volume} {36}},\ \bibinfo
		{pages} {3641} (\bibinfo {year} {1987})}\BibitemShut {NoStop}%
	\bibitem [{\citenamefont {Halliwell}(1988)}]{Halliwell_ordering}%
	\BibitemOpen
	\bibfield  {author} {\bibinfo {author} {\bibfnamefont {J.~J.}\ \bibnamefont
			{Halliwell}},\ }\bibfield  {title} {\bibinfo {title} {Derivation of the
			wheeler-dewitt equation from a path integral for minisuperspace models},\
	}\href {https://doi.org/10.1103/PhysRevD.38.2468} {\bibfield  {journal}
		{\bibinfo  {journal} {Phys. Rev. D}\ }\textbf {\bibinfo {volume} {38}},\
		\bibinfo {pages} {2468} (\bibinfo {year} {1988})}\BibitemShut {NoStop}%
	\bibitem [{\citenamefont {Hawking}\ and\ \citenamefont
		{Page}(1986)}]{Hawking:1985bk}%
	\BibitemOpen
	\bibfield  {author} {\bibinfo {author} {\bibfnamefont {S.~W.}\ \bibnamefont
			{Hawking}}\ and\ \bibinfo {author} {\bibfnamefont {D.~N.}\ \bibnamefont
			{Page}},\ }\bibfield  {title} {\bibinfo {title} {{Operator Ordering and the
				Flatness of the Universe}},\ }\href
	{https://doi.org/10.1016/0550-3213(86)90478-5} {\bibfield  {journal}
		{\bibinfo  {journal} {Nucl. Phys. B}\ }\textbf {\bibinfo {volume} {264}},\
		\bibinfo {pages} {185} (\bibinfo {year} {1986})}\BibitemShut {NoStop}%
	\bibitem [{\citenamefont {Vilenkin}(1988{\natexlab{b}})}]{Vilenkin_ordering}%
	\BibitemOpen
	\bibfield  {author} {\bibinfo {author} {\bibfnamefont {A.}~\bibnamefont
			{Vilenkin}},\ }\bibfield  {title} {\bibinfo {title} {Quantum cosmology and
			the initial state of the universe},\ }\href
	{https://doi.org/10.1103/PhysRevD.37.888} {\bibfield  {journal} {\bibinfo
			{journal} {Phys. Rev. D}\ }\textbf {\bibinfo {volume} {37}},\ \bibinfo
		{pages} {888} (\bibinfo {year} {1988}{\natexlab{b}})}\BibitemShut {NoStop}%
	\bibitem [{\citenamefont {Louko}(1988)}]{Louko:1988zb}%
	\BibitemOpen
	\bibfield  {author} {\bibinfo {author} {\bibfnamefont {J.}~\bibnamefont
			{Louko}},\ }\bibfield  {title} {\bibinfo {title} {{Semiclassical Path Measure
				and Factor Ordering in Quantum Cosmology}},\ }\href
	{https://doi.org/10.1016/0003-4916(88)90170-4} {\bibfield  {journal}
		{\bibinfo  {journal} {Annals Phys.}\ }\textbf {\bibinfo {volume} {181}},\
		\bibinfo {pages} {318} (\bibinfo {year} {1988})}\BibitemShut {NoStop}%
        \bibitem{Halliwell}
        J.~J. Halliwell, ``Derivation of the wheeler-dewitt equation from a path
  integral for minisuperspace models,''
  \href{http://dx.doi.org/10.1103/PhysRevD.38.2468}{{\em Phys. Rev. D}
  {\bfseries 38} (Oct, 1988) 2468--2481}.%
	\bibitem [{\citenamefont {Bojowald}(2002)}]{Bojowald:2002gz}%
	\BibitemOpen
	\bibfield  {author} {\bibinfo {author} {\bibfnamefont {M.}~\bibnamefont
			{Bojowald}},\ }\bibfield  {title} {\bibinfo {title} {{Isotropic loop quantum
				cosmology}},\ }\href {https://doi.org/10.1088/0264-9381/19/10/313} {\bibfield
		{journal} {\bibinfo  {journal} {Class. Quant. Grav.}\ }\textbf {\bibinfo
			{volume} {19}},\ \bibinfo {pages} {2717} (\bibinfo {year} {2002})}\ \BibitemShut
	{NoStop}%
        \bibitem{Kontoleon:1998pw}
N.~Kontoleon and D.~L. Wiltshire, ``{Operator ordering and consistency of the
  wave function of the universe},''
  \href{http://dx.doi.org/10.1103/PhysRevD.59.063513}{{\em Phys. Rev. D}
  {\bfseries 59} (1999) 063513},
  \href{http://arxiv.org/abs/gr-qc/9807075}{{\ttfamily arXiv:gr-qc/9807075}}.
	\bibitem [{\citenamefont {Bojowald}\ and\ \citenamefont
		{Simpson}(2014)}]{Bojowald:2014ija}%
	\BibitemOpen
	\bibfield  {author} {\bibinfo {author} {\bibfnamefont {M.}~\bibnamefont
			{Bojowald}}\ and\ \bibinfo {author} {\bibfnamefont {D.}~\bibnamefont
			{Simpson}},\ }\bibfield  {title} {\bibinfo {title} {{Factor ordering and
				large-volume dynamics in quantum cosmology}},\ }\href
	{https://doi.org/10.1088/0264-9381/31/18/185016} {\bibfield  {journal}
		{\bibinfo  {journal} {Class. Quant. Grav.}\ }\textbf {\bibinfo {volume}
			{31}},\ \bibinfo {pages} {185016} (\bibinfo {year} {2014})}\ \BibitemShut
	{NoStop}%
 \bibitem [{\citenamefont {Rostami}\ \emph {et~al.}(2015)\citenamefont
		{Rostami}, \citenamefont {Jalalzadeh},\ and\ \citenamefont
		{Moniz}}]{Rostami_ordering}%
	\BibitemOpen
	\bibfield  {author} {\bibinfo {author} {\bibfnamefont {T.}~\bibnamefont
			{Rostami}}, \bibinfo {author} {\bibfnamefont {S.}~\bibnamefont
			{Jalalzadeh}},\ and\ \bibinfo {author} {\bibfnamefont {P.~V.}\ \bibnamefont
			{Moniz}},\ }\bibfield  {title} {\bibinfo {title} {Quantum cosmological
			intertwining: Factor ordering and boundary conditions from hidden
			symmetries},\ }\href {https://doi.org/10.1103/PhysRevD.92.023526} {\bibfield
		{journal} {\bibinfo  {journal} {Phys. Rev. D}\ }\textbf {\bibinfo {volume}
			{92}},\ \bibinfo {pages} {023526} (\bibinfo {year} {2015})}\BibitemShut
	{NoStop}%
	\bibitem [{\citenamefont {Kiefer}\ and\ \citenamefont
		{Schmitz}(2019)}]{kiefer_singularity_2019}%
	\BibitemOpen
	\bibfield  {author} {\bibinfo {author} {\bibfnamefont {C.}~\bibnamefont
			{Kiefer}}\ and\ \bibinfo {author} {\bibfnamefont {T.}~\bibnamefont
			{Schmitz}},\ }\bibfield  {title} {\bibinfo {title} {Singularity avoidance for
			collapsing quantum dust in the {Lema{\^i}tre}-{Tolman}-{Bondi} model},\
	}\href {https://doi.org/10.1103/PhysRevD.99.126010} {\bibfield  {journal}
		{\bibinfo  {journal} {Physical Review D}\ }\textbf {\bibinfo {volume} {99}},\
		\bibinfo {pages} {126010} (\bibinfo {year} {2019}).}\BibitemShut {NoStop}%
    \bibitem{Matsui:2021yte}
H.~Matsui, S.~Mukohyama, and A.~Naruko, ``{DeWitt boundary condition is
  consistent in Ho\v{r}ava-Lifshitz quantum gravity},''
  \href{http://dx.doi.org/10.1016/j.physletb.2022.137340}{{\em Phys. Lett. B}
  {\bfseries 833} (2022) 137340},
  \href{http://arxiv.org/abs/2111.00665}{{\ttfamily arXiv:2111.00665 [gr-qc]}}.
  \bibitem [{\citenamefont {Sahota}\ and\ \citenamefont
		{Lochan}(2021)}]{Sahota_Infrared}%
	\BibitemOpen
	\bibfield  {author} {\bibinfo {author} {\bibfnamefont {H.~S.}\ \bibnamefont
			{Sahota}}\ and\ \bibinfo {author} {\bibfnamefont {K.}~\bibnamefont
			{Lochan}},\ }\bibfield  {title} {\bibinfo {title} {Infrared signatures of a
			quantum bounce in a minisuperspace analysis of lema\^{\i}tre-tolman-bondi
			dust collapse},\ }\href {https://doi.org/10.1103/PhysRevD.104.126027}
	{\bibfield  {journal} {\bibinfo  {journal} {Phys. Rev. D}\ }\textbf {\bibinfo
			{volume} {104}},\ \bibinfo {pages} {126027} (\bibinfo {year}
		{2021})}\BibitemShut {NoStop}%
	\bibitem [{\citenamefont {Sahota}\ and\ \citenamefont {Lochan}(2023)}]{Sahota2023}%
	\BibitemOpen
	\bibfield  {author} {\bibinfo {author} {\bibfnamefont {H.~S.}\ \bibnamefont {Sahota}}\ and\ \bibinfo {author} {\bibfnamefont {K.}~\bibnamefont {Lochan}},\ }\bibfield  {title} {\enquote {\bibinfo {title} {Analyzing quantum gravity spillover in the semiclassical regime},}\ }\href {\doibase 10.1140/epjc/s10052-023-12311-2} {\bibfield  {journal} {\bibinfo  {journal} {The European Physical Journal C}\ }\textbf {\bibinfo {volume} {83}},\ \bibinfo {pages} {1162} (\bibinfo {year} {2023})}\ \BibitemShut {NoStop}%
	\bibitem [{\citenamefont {Thebault}(2023)}]{Thebault:2022dmv}%
	\BibitemOpen
	\bibfield  {author} {\bibinfo {author} {\bibfnamefont {K.~P.~Y.}\
			\bibnamefont {Thebault}},\ }\bibfield  {title} {\bibinfo {title} {{Big bang
				singularity resolution in quantum cosmology}},\ }\href
	{https://doi.org/10.1088/1361-6382/acb752} {\bibfield  {journal} {\bibinfo
			{journal} {Class. Quant. Grav.}\ }\textbf {\bibinfo {volume} {40}},\ \bibinfo
		{pages} {055007} (\bibinfo {year} {2023})}\ \BibitemShut
	{NoStop}%
	\bibitem [{\citenamefont {Bojowald}(2007)}]{Bojowald:2007}%
	\BibitemOpen
	\bibfield  {author} {\bibinfo {author} {\bibfnamefont {M.}~\bibnamefont
			{Bojowald}},\ }\bibfield  {title} {\bibinfo {title} {Dynamical coherent
			states and physical solutions of quantum cosmological bounces},\ }\href
	{https://doi.org/10.1103/PhysRevD.75.123512} {\bibfield  {journal} {\bibinfo
			{journal} {Phys. Rev. D}\ }\textbf {\bibinfo {volume} {75}},\ \bibinfo
		{pages} {123512} (\bibinfo {year} {2007})}\BibitemShut {NoStop}%
	\bibitem [{\citenamefont {Ashtekar}\ \emph {et~al.}(2009)\citenamefont
		{Ashtekar}, \citenamefont {Kaminski},\ and\ \citenamefont
		{Lewandowski}}]{Ashtekar:2009mb}%
	\BibitemOpen
	\bibfield  {author} {\bibinfo {author} {\bibfnamefont {A.}~\bibnamefont
			{Ashtekar}}, \bibinfo {author} {\bibfnamefont {W.}~\bibnamefont {Kaminski}},\
		and\ \bibinfo {author} {\bibfnamefont {J.}~\bibnamefont {Lewandowski}},\
	}\bibfield  {title} {\bibinfo {title} {{Quantum field theory on a
				cosmological, quantum space-time}},\ }\href
	{https://doi.org/10.1103/PhysRevD.79.064030} {\bibfield  {journal} {\bibinfo
			{journal} {Phys. Rev. D}\ }\textbf {\bibinfo {volume} {79}},\ \bibinfo
		{pages} {064030} (\bibinfo {year} {2009})}\ \BibitemShut
	{NoStop}%
	\bibitem [{\citenamefont {Agullo}\ \emph
		{et~al.}(2013{\natexlab{a}})\citenamefont {Agullo}, \citenamefont
		{Ashtekar},\ and\ \citenamefont {Nelson}}]{Agullo:2012fc}%
	\BibitemOpen
	\bibfield  {author} {\bibinfo {author} {\bibfnamefont {I.}~\bibnamefont
			{Agullo}}, \bibinfo {author} {\bibfnamefont {A.}~\bibnamefont {Ashtekar}},\
		and\ \bibinfo {author} {\bibfnamefont {W.}~\bibnamefont {Nelson}},\
	}\bibfield  {title} {\bibinfo {title} {{Extension of the quantum theory of
				cosmological perturbations to the Planck era}},\ }\href
	{https://doi.org/10.1103/PhysRevD.87.043507} {\bibfield  {journal} {\bibinfo
			{journal} {Phys. Rev. D}\ }\textbf {\bibinfo {volume} {87}},\ \bibinfo
		{pages} {043507} (\bibinfo {year} {2013}{\natexlab{a}})}\ \BibitemShut
	{NoStop}%
	\bibitem [{\citenamefont {Agullo}\ \emph {et~al.}(2012)\citenamefont {Agullo},
		\citenamefont {Ashtekar},\ and\ \citenamefont {Nelson}}]{Agullo:2012sh}%
	\BibitemOpen
	\bibfield  {author} {\bibinfo {author} {\bibfnamefont {I.}~\bibnamefont
			{Agullo}}, \bibinfo {author} {\bibfnamefont {A.}~\bibnamefont {Ashtekar}},\
		and\ \bibinfo {author} {\bibfnamefont {W.}~\bibnamefont {Nelson}},\
	}\bibfield  {title} {\bibinfo {title} {{A Quantum Gravity Extension of the
				Inflationary Scenario}},\ }\href
	{https://doi.org/10.1103/PhysRevLett.109.251301} {\bibfield  {journal}
		{\bibinfo  {journal} {Phys. Rev. Lett.}\ }\textbf {\bibinfo {volume} {109}},\
		\bibinfo {pages} {251301} (\bibinfo {year} {2012})}\ \BibitemShut
	{NoStop}%
	\bibitem [{\citenamefont {Agullo}\ \emph
		{et~al.}(2013{\natexlab{b}})\citenamefont {Agullo}, \citenamefont
		{Ashtekar},\ and\ \citenamefont {Nelson}}]{Agullo:2013ai}%
	\BibitemOpen
	\bibfield  {author} {\bibinfo {author} {\bibfnamefont {I.}~\bibnamefont
			{Agullo}}, \bibinfo {author} {\bibfnamefont {A.}~\bibnamefont {Ashtekar}},\
		and\ \bibinfo {author} {\bibfnamefont {W.}~\bibnamefont {Nelson}},\
	}\bibfield  {title} {\bibinfo {title} {{The pre-inflationary dynamics of loop
				quantum cosmology: Confronting quantum gravity with observations}},\ }\href
	{https://doi.org/10.1088/0264-9381/30/8/085014} {\bibfield  {journal}
		{\bibinfo  {journal} {Class. Quant. Grav.}\ }\textbf {\bibinfo {volume}
			{30}},\ \bibinfo {pages} {085014} (\bibinfo {year} {2013}{\natexlab{b}})}\
	\BibitemShut {NoStop}%
	\bibitem [{\citenamefont {Schutz}(1970)}]{Schutz_1970}%
	\BibitemOpen
	\bibfield  {author} {\bibinfo {author} {\bibfnamefont {B.~F.}\ \bibnamefont
			{Schutz}},\ }\bibfield  {title} {\bibinfo {title} {Perfect fluids in general
			relativity: Velocity potentials and a variational principle},\ }\href
	{https://doi.org/10.1103/PhysRevD.2.2762} {\bibfield  {journal} {\bibinfo
			{journal} {Phys. Rev. D}\ }\textbf {\bibinfo {volume} {2}},\ \bibinfo {pages}
		{2762} (\bibinfo {year} {1970})}\BibitemShut {NoStop}%
	\bibitem [{\citenamefont {Schutz}(1971)}]{Schutz_1971}%
	\BibitemOpen
	\bibfield  {author} {\bibinfo {author} {\bibfnamefont {B.~F.}\ \bibnamefont
			{Schutz}},\ }\bibfield  {title} {\bibinfo {title} {Hamiltonian theory of a
			relativistic perfect fluid},\ }\href
	{https://doi.org/10.1103/PhysRevD.4.3559} {\bibfield  {journal} {\bibinfo
			{journal} {Phys. Rev. D}\ }\textbf {\bibinfo {volume} {4}},\ \bibinfo {pages}
		{3559} (\bibinfo {year} {1971})}\BibitemShut {NoStop}%
  \bibitem [{\citenamefont {Peter}\ and\ \citenamefont
			{Pinto-Neto}(2008)}]{Peter:2008qz}%
		\BibitemOpen
		\bibfield  {author} {\bibinfo {author} {\bibfnamefont {P.}~\bibnamefont
				{Peter}}\ and\ \bibinfo {author} {\bibfnamefont {N.}~\bibnamefont
				{Pinto-Neto}},\ }\bibfield  {title} {\bibinfo {title} {Cosmology without inflation},\ }\href {\doibase 10.1103/PhysRevD.78.063506} {\bibfield
			{journal} {\bibinfo  {journal} {Phys. Rev. D}\ }\textbf {\bibinfo {volume}
				{78}},\ \bibinfo {pages} {063506} (\bibinfo {year} {2008})}\ \BibitemShut
		{NoStop}%
  \bibitem{ArfkenWeber} G.~Arfken and H.~Weber, {\em Mathematical Methods for Physicists}, 6th ed. (Academic Press, Boston, 2005).
	\bibitem [{\citenamefont {H{\'a}j{\'i}{\v c}ek}\ and\ \citenamefont
		{Kiefer}(2001)}]{hajicek_singularity_2001}%
	\BibitemOpen
	\bibfield  {author} {\bibinfo {author} {\bibfnamefont {P.}~\bibnamefont
			{H{\'a}j{\'i}{\v c}ek}}\ and\ \bibinfo {author} {\bibfnamefont
			{C.}~\bibnamefont {Kiefer}},\ }\bibfield  {title} {\bibinfo {title}
		{{Singularity} {avoidance} {by} {collapsing} {shells} {in} {quantum}
			{gravity}},\ }\href {https://doi.org/10.1142/S0218271801001578} {\bibfield
		{journal} {\bibinfo  {journal} {International Journal of Modern Physics D}\
		}\textbf {\bibinfo {volume} {10}},\ \bibinfo {pages} {775} (\bibinfo {year}
		{2001})}\BibitemShut {NoStop}%
             \bibitem [{\citenamefont {BALLENTINE}(1970)}]{Ballentine}%
  \BibitemOpen
  \bibfield  {author} {\bibinfo {author} {\bibfnamefont {L.~E.}\ \bibnamefont
  {Ballentine}},\ }\bibfield  {title} {\bibinfo {title}
		{The Statistical Interpretation of Quantum Mechanics},\ }\href {\doibase 10.1103/RevModPhys.42.358} {\bibfield
  {journal} {\bibinfo  {journal} {Rev. Mod. Phys.}\ }\textbf {\bibinfo {volume}
  {42}},\ \bibinfo {pages} {358} (\bibinfo {year} {1970})}\BibitemShut
  {NoStop}%
	\bibitem [{\citenamefont {GRADSHTEYN}\ and\ \citenamefont
		{RYZHIK}(1980)}]{GRADSHTEYN1980635}%
	\BibitemOpen
	\bibfield  {author} {\bibinfo {author} {\bibfnamefont {I.}~\bibnamefont
			{Gradshteyn}}\ and\ \bibinfo {author} {\bibfnamefont {I.}~\bibnamefont
			{Ryzhik}},\ }\bibfield  {title} {\bibinfo {title} {6.-7 - definite integrals
			of special functions},\ }in\ \href
	{https://doi.org/https://doi.org/10.1016/B978-0-12-294760-5.50019-2} {\emph
		{\bibinfo {booktitle} {Table of Integrals, Series, and Products}}},\ \bibinfo
	{editor} {edited by\ \bibinfo {editor} {\bibfnamefont {I.}~\bibnamefont
			{Gradshteyn}}\ and\ \bibinfo {editor} {\bibfnamefont {I.}~\bibnamefont
			{Ryzhik}}}\ (\bibinfo  {publisher} {Academic Press},\ \bibinfo {year}
	{1980})\ pp.\ \bibinfo {pages} {635--903}\BibitemShut {NoStop}%
	\bibitem [{\citenamefont {Andrews}(1998)}]{Andrews_WP}%
	\BibitemOpen
	\bibfield  {author} {\bibinfo {author} {\bibfnamefont {M.}~\bibnamefont
			{Andrews}},\ }\bibfield  {title} {\bibinfo {title} {{Wave packets bouncing
				off walls}},\ }\href {https://doi.org/10.1119/1.18854} {\bibfield  {journal}
		{\bibinfo  {journal} {American Journal of Physics}\ }\textbf {\bibinfo
			{volume} {66}},\ \bibinfo {pages} {252} (\bibinfo {year} {1998})}\
	\BibitemShut {NoStop}%
	\bibitem [{\citenamefont {Robinett}(2004)}]{ROBINETT20041}%
	\BibitemOpen
	\bibfield  {author} {\bibinfo {author} {\bibfnamefont {R.}~\bibnamefont
			{Robinett}},\ }\bibfield  {title} {\bibinfo {title} {Quantum wave packet
			revivals},\ }\href
	{https://doi.org/https://doi.org/10.1016/j.physrep.2003.11.002} {\bibfield
		{journal} {\bibinfo  {journal} {Physics Reports}\ }\textbf {\bibinfo {volume}
			{392}},\ \bibinfo {pages} {1} (\bibinfo {year} {2004})}\BibitemShut {NoStop}%
	\bibitem [{\citenamefont {Alexandre}\ and\ \citenamefont
		{Magueijo}(2022)}]{Alexandre:2022ijm}%
	\BibitemOpen
	\bibfield  {author} {\bibinfo {author} {\bibfnamefont {B.}~\bibnamefont
			{Alexandre}}\ and\ \bibinfo {author} {\bibfnamefont {J.}~\bibnamefont
			{Magueijo}},\ }\bibfield  {title} {\bibinfo {title} {{Possible quantum
				effects at the transition from cosmological deceleration to acceleration}},\
	}\href {https://doi.org/10.1103/PhysRevD.106.063520} {\bibfield  {journal}
		{\bibinfo  {journal} {Phys. Rev. D}\ }\textbf {\bibinfo {volume} {106}},\
		\bibinfo {pages} {063520} (\bibinfo {year} {2022})}\ \BibitemShut
	{NoStop}%
	\bibitem [{\citenamefont {Alexandre}\ and\ \citenamefont
		{Magueijo}(2023)}]{Alexandre:2022npo}%
	\BibitemOpen
	\bibfield  {author} {\bibinfo {author} {\bibfnamefont {B.}~\bibnamefont
			{Alexandre}}\ and\ \bibinfo {author} {\bibfnamefont {J.}~\bibnamefont
			{Magueijo}},\ }\bibfield  {title} {\bibinfo {title} {{Unimodular
				Hartle-Hawking wave packets and their probability interpretation}},\ }\href
	{https://doi.org/10.1103/PhysRevD.107.063501} {\bibfield  {journal} {\bibinfo
			{journal} {Phys. Rev. D}\ }\textbf {\bibinfo {volume} {107}},\ \bibinfo
		{pages} {063501} (\bibinfo {year} {2023})}\ \BibitemShut
	{NoStop}%
	\bibitem [{\citenamefont {Gielen}\ and\ \citenamefont
		{Magueijo}(2023)}]{Gielen:2022dhg}%
	\BibitemOpen
	\bibfield  {author} {\bibinfo {author} {\bibfnamefont {S.}~\bibnamefont
			{Gielen}}\ and\ \bibinfo {author} {\bibfnamefont {J.}~\bibnamefont
			{Magueijo}},\ }\bibfield  {title} {\bibinfo {title} {{Quantum analysis of the
				recent cosmological bounce in the comoving Hubble length}},\ }\href
	{https://doi.org/10.1103/PhysRevD.107.023518} {\bibfield  {journal} {\bibinfo
			{journal} {Phys. Rev. D}\ }\textbf {\bibinfo {volume} {107}},\ \bibinfo
		{pages} {023518} (\bibinfo {year} {2023})}\ \BibitemShut
	{NoStop}%
        \bibitem [{\citenamefont {Gotay}\ and\ \citenamefont {Demaret}(1983)}]{Gotay_Singularity}%
  \BibitemOpen
  \bibfield  {author} {\bibinfo {author} {\bibfnamefont {M.~J.}\ \bibnamefont {Gotay}}\ and\ \bibinfo {author} {\bibfnamefont {J.}~\bibnamefont {Demaret}},\ }\bibfield  {title} {\enquote {\bibinfo {title} {Quantum cosmological singularities},}\ }\href {\doibase 10.1103/PhysRevD.28.2402} {\bibfield  {journal} {\bibinfo  {journal} {Phys. Rev. D}\ }\textbf {\bibinfo {volume} {28}},\ \bibinfo {pages} {2402--2413} (\bibinfo {year} {1983})}\BibitemShut {NoStop}%
\bibitem [{\citenamefont {Ma\l{}kiewicz}\ \emph {et~al.}(2020)\citenamefont {Ma\l{}kiewicz}, \citenamefont {Peter},\ and\ \citenamefont {Vitenti}}]{Malkiewicz:2019azw}%
  \BibitemOpen
  \bibfield  {author} {\bibinfo {author} {\bibfnamefont {P.}~\bibnamefont {Ma\l{}kiewicz}}, \bibinfo {author} {\bibfnamefont {P.}~\bibnamefont {Peter}}, \ and\ \bibinfo {author} {\bibfnamefont {S.~D.~P.}\ \bibnamefont {Vitenti}},\ }\bibfield  {title} {\enquote {\bibinfo {title} {{Quantum empty Bianchi I spacetime with internal time}},}\ }\href {\doibase 10.1103/PhysRevD.101.046012} {\bibfield  {journal} {\bibinfo  {journal} {Phys. Rev. D}\ }\textbf {\bibinfo {volume} {101}},\ \bibinfo {pages} {046012} (\bibinfo {year} {2020})}\ \BibitemShut {NoStop}%
\bibitem [{\citenamefont {Gielen}\ and\ \citenamefont {Men\'endez-Pidal}(2020)}]{Gielen:2020abd}%
  \BibitemOpen
  \bibfield  {author} {\bibinfo {author} {\bibfnamefont {S.}~\bibnamefont {Gielen}}\ and\ \bibinfo {author} {\bibfnamefont {L.}~\bibnamefont {Men\'endez-Pidal}},\ }\bibfield  {title} {\enquote {\bibinfo {title} {{Singularity resolution depends on the clock}},}\ }\href {\doibase 10.1088/1361-6382/abb14f} {\bibfield  {journal} {\bibinfo  {journal} {Class. Quant. Grav.}\ }\textbf {\bibinfo {volume} {37}},\ \bibinfo {pages} {205018} (\bibinfo {year} {2020})}\ \BibitemShut {NoStop}%
\bibitem [{\citenamefont {Gielen}\ and\ \citenamefont {Men\'endez-Pidal}(2022{\natexlab{a}})}]{Gielen:2021igw}%
  \BibitemOpen
  \bibfield  {author} {\bibinfo {author} {\bibfnamefont {S.}~\bibnamefont {Gielen}}\ and\ \bibinfo {author} {\bibfnamefont {L.}~\bibnamefont {Men\'endez-Pidal}},\ }\bibfield  {title} {\enquote {\bibinfo {title} {{Unitarity, clock dependence and quantum recollapse in quantum cosmology}},}\ }\href {\doibase 10.1088/1361-6382/ac504f} {\bibfield  {journal} {\bibinfo  {journal} {Class. Quant. Grav.}\ }\textbf {\bibinfo {volume} {39}},\ \bibinfo {pages} {075011} (\bibinfo {year} {2022}{\natexlab{a}})}\ \BibitemShut {NoStop}%
	\bibitem [{\citenamefont {Alvarenga}\ and\ \citenamefont
		{Lemos}(1998)}]{Alvarenga:1998wx}%
	\BibitemOpen
	\bibfield  {author} {\bibinfo {author} {\bibfnamefont {F.~G.}\ \bibnamefont
			{Alvarenga}}\ and\ \bibinfo {author} {\bibfnamefont {N.~A.}\ \bibnamefont
			{Lemos}},\ }\bibfield  {title} {\bibinfo {title} {{Dynamical vacuum in
				quantum cosmology}},\ }\href {https://doi.org/10.1023/A:1018896900336}
	{\bibfield  {journal} {\bibinfo  {journal} {Gen. Rel. Grav.}\ }\textbf
		{\bibinfo {volume} {30}},\ \bibinfo {pages} {681} (\bibinfo {year} {1998})}\
	\BibitemShut {NoStop}%
	\bibitem [{\citenamefont {Alvarenga}\ \emph {et~al.}(2002)\citenamefont
		{Alvarenga}, \citenamefont {Fabris}, \citenamefont {Lemos},\ and\
		\citenamefont {Monerat}}]{Alvarenga:2001nm}%
	\BibitemOpen
	\bibfield  {author} {\bibinfo {author} {\bibfnamefont {F.~G.}\ \bibnamefont
			{Alvarenga}}, \bibinfo {author} {\bibfnamefont {J.~C.}\ \bibnamefont
			{Fabris}}, \bibinfo {author} {\bibfnamefont {N.~A.}\ \bibnamefont {Lemos}},\
		and\ \bibinfo {author} {\bibfnamefont {G.~A.}\ \bibnamefont {Monerat}},\
	}\bibfield  {title} {\bibinfo {title} {{Quantum cosmological perfect fluid
				models}},\ }\href {https://doi.org/10.1023/A:1015986011295} {\bibfield
		{journal} {\bibinfo  {journal} {Gen. Rel. Grav.}\ }\textbf {\bibinfo {volume}
			{34}},\ \bibinfo {pages} {651} (\bibinfo {year} {2002})}\ \BibitemShut
	{NoStop}%
	\bibitem [{\citenamefont {Batista}\ \emph {et~al.}(2002)\citenamefont
		{Batista}, \citenamefont {Fabris}, \citenamefont {Goncalves},\ and\
		\citenamefont {Tossa}}]{Batista:2001ti}%
	\BibitemOpen
	\bibfield  {author} {\bibinfo {author} {\bibfnamefont {A.~B.}\ \bibnamefont
			{Batista}}, \bibinfo {author} {\bibfnamefont {J.~C.}\ \bibnamefont {Fabris}},
		\bibinfo {author} {\bibfnamefont {S.~V.~B.}\ \bibnamefont {Goncalves}},\ and\
		\bibinfo {author} {\bibfnamefont {J.}~\bibnamefont {Tossa}},\ }\bibfield
	{title} {\bibinfo {title} {{Quantum cosmological perfect fluid model and its
				classical analog}},\ }\href {https://doi.org/10.1103/PhysRevD.65.063519}
	{\bibfield  {journal} {\bibinfo  {journal} {Phys. Rev. D}\ }\textbf {\bibinfo
			{volume} {65}},\ \bibinfo {pages} {063519} (\bibinfo {year} {2002})}\
	\BibitemShut {NoStop}%
	\bibitem [{\citenamefont {Pedram}\ \emph
		{et~al.}(2007{\natexlab{a}})\citenamefont {Pedram}, \citenamefont
		{Jalalzadeh},\ and\ \citenamefont {Gousheh}}]{Pedram:2007iv}%
	\BibitemOpen
	\bibfield  {author} {\bibinfo {author} {\bibfnamefont {P.}~\bibnamefont
			{Pedram}}, \bibinfo {author} {\bibfnamefont {S.}~\bibnamefont {Jalalzadeh}},\
		and\ \bibinfo {author} {\bibfnamefont {S.~S.}\ \bibnamefont {Gousheh}},\
	}\bibfield  {title} {\bibinfo {title} {{Schrodinger-Wheeler-DeWitt equation
				in chaplygin gas FRW cosmological model}},\ }\href
	{https://doi.org/10.1007/s10773-007-9436-9} {\bibfield  {journal} {\bibinfo
			{journal} {Int. J. Theor. Phys.}\ }\textbf {\bibinfo {volume} {46}},\
		\bibinfo {pages} {3201} (\bibinfo {year} {2007}{\natexlab{a}})}\ \BibitemShut
	{NoStop}%
	\bibitem [{\citenamefont {Pedram}\ \emph
		{et~al.}(2007{\natexlab{b}})\citenamefont {Pedram}, \citenamefont
		{Jalalzadeh},\ and\ \citenamefont {Gousheh}}]{Pedram:2007ck}%
	\BibitemOpen
	\bibfield  {author} {\bibinfo {author} {\bibfnamefont {P.}~\bibnamefont
			{Pedram}}, \bibinfo {author} {\bibfnamefont {S.}~\bibnamefont {Jalalzadeh}},\
		and\ \bibinfo {author} {\bibfnamefont {S.~S.}\ \bibnamefont {Gousheh}},\
	}\bibfield  {title} {\bibinfo {title} {{Quantum Stephani exact cosmological
				solutions and the selection of time variable}},\ }\href
	{https://doi.org/10.1088/0264-9381/24/22/014} {\bibfield  {journal} {\bibinfo
			{journal} {Class. Quant. Grav.}\ }\textbf {\bibinfo {volume} {24}},\
		\bibinfo {pages} {5515} (\bibinfo {year} {2007}{\natexlab{b}})}\ \BibitemShut
	{NoStop}%
	\bibitem [{\citenamefont {Pedram}\ and\ \citenamefont
		{Jalalzadeh}(2008)}]{Pedram:2007ud}%
	\BibitemOpen
	\bibfield  {author} {\bibinfo {author} {\bibfnamefont {P.}~\bibnamefont
			{Pedram}}\ and\ \bibinfo {author} {\bibfnamefont {S.}~\bibnamefont
			{Jalalzadeh}},\ }\bibfield  {title} {\bibinfo {title} {{Quantum FRW
				cosmological solutions in the presence of Chaplygin gas and perfect fluid}},\
	}\href {https://doi.org/10.1016/j.physletb.2007.11.013} {\bibfield  {journal}
		{\bibinfo  {journal} {Phys. Lett. B}\ }\textbf {\bibinfo {volume} {659}},\
		\bibinfo {pages} {6} (\bibinfo {year} {2008})}\ \BibitemShut
	{NoStop}%
	\bibitem [{\citenamefont {Pedram}\ \emph {et~al.}(2008)\citenamefont {Pedram},
		\citenamefont {Mirzaei}, \citenamefont {Jalalzadeh},\ and\ \citenamefont
		{Gousheh}}]{Pedram:2007fm}%
	\BibitemOpen
	\bibfield  {author} {\bibinfo {author} {\bibfnamefont {P.}~\bibnamefont
			{Pedram}}, \bibinfo {author} {\bibfnamefont {M.}~\bibnamefont {Mirzaei}},
		\bibinfo {author} {\bibfnamefont {S.}~\bibnamefont {Jalalzadeh}},\ and\
		\bibinfo {author} {\bibfnamefont {S.~S.}\ \bibnamefont {Gousheh}},\
	}\bibfield  {title} {\bibinfo {title} {{Perfect fluid quantum Universe in the
				presence of negative cosmological constant}},\ }\href
	{https://doi.org/10.1007/s10714-007-0566-4} {\bibfield  {journal} {\bibinfo
			{journal} {Gen. Rel. Grav.}\ }\textbf {\bibinfo {volume} {40}},\ \bibinfo
		{pages} {1663} (\bibinfo {year} {2008})}\ \BibitemShut
	{NoStop}%
        \bibitem [{\citenamefont {de~la Madrid}(2005)}]{Madrid_2005}%
  \BibitemOpen
  \bibfield  {author} {\bibinfo {author} {\bibfnamefont {R.}~\bibnamefont {de~la Madrid}},\ }\bibfield  {title} {\enquote {\bibinfo {title} {The role of the rigged hilbert space in quantum mechanics},}\ }\href {\doibase 10.1088/0143-0807/26/2/008} {\bibfield  {journal} {\bibinfo  {journal} {European Journal of Physics}\ }\textbf {\bibinfo {volume} {26}},\ \bibinfo {pages} {287--312} (\bibinfo {year} {2005})}\BibitemShut {NoStop}%
\end{thebibliography}
\end{document}